\documentclass[12pt]{article}
\usepackage{epsfig}
\def\be{\begin{equation}}
\def\ee{\end{equation}}
\def\bc{\begin{center}}
\def\ec{\end{center}}
\def\bea{\begin{eqnarray}}
\def\eea{\end{eqnarray}}

\def\nn{\nonumber}
\def\ad{\dot{\alpha}}
\def\cP{{\cal P}}

\def\U{{\cal U}}

\catcode`@=11
\def\marginnote#1{}
\newcount\hour
\newcount\minute
\newtoks\amorpm
\hour=\time\divide\hour by60
\minute=\time{\multiply\hour by60 \global\advance\minute by-\hour}
\edef\standardtime{{\ifnum\hour<12 \global\amorpm={am}%
        \else\global\amorpm={pm}\advance\hour by-12 \fi
        \ifnum\hour=0 \hour=12 \fi
        \number\hour:\ifnum\minute<10 0\fi\number\minute\the\amorpm}}
\edef\militarytime{\number\hour:\ifnum\minute<10 0\fi\number\minute}
\def\draftlabel#1{{\@bsphack\if@filesw {\let\thepage\relax
   \xdef\@gtempa{\write\@auxout{\string
      \newlabel{#1}{{\@currentlabel}{\thepage}}}}}\@gtempa
   \if@nobreak \ifvmode\nobreak\fi\fi\fi\@esphack}
        \gdef\@eqnlabel{#1}}
\def\@eqnlabel{}
\def\@vacuum{}
\def\draftmarginnote#1{\marginpar{\raggedright\scriptsize\tt#1}}
\def\draft{\oddsidemargin 0.0truein
        \def\@oddfoot{\sl preliminary draft \hfil
        \rm\thepage\hfil\sl\today\quad\militarytime}
        \let\@evenfoot\@oddfoot \overfullrule 3pt
        \let\label=\draftlabel
        \let\marginnote=\draftmarginnote
   \def\@eqnnum{(\theequation)\rlap{\kern\marginparsep\tt\@eqnlabel}%
\global\let\@eqnlabel\@vacuum}  }
\catcode`@=12
\begin{document}
\begin{titlepage}
\vspace*{-1cm}

\begin{flushright}
UNIL-IPT-01-15\\
hep-th/0110022
\end{flushright}

\vskip 2.0cm

\begin{center}
{\Large\bf Gravity-Mediated Supersymmetry Breaking\\
\vskip 0.1cm
in the Brane World}
\end{center}
\vskip 1.0cm

\begin{center}
{\sc Tony Gherghetta$^a$~\footnote{E-mail: tony.gherghetta@cern.ch} and
Antonio Riotto$^b$~\footnote{E-mail: antonio.riotto@pd.infn.it}}\\
\vskip .9cm
{\it $^a$ Institute of Theoretical Physics, University of Lausanne, \\
CH-1015 Lausanne, Switzerland}\\
{\it $^b$ INFN, Sezione di Padova, Via Marzolo~8, I-35131 Padua, Italy}
\end{center}

\vskip 1.5cm

\begin{abstract}
\noindent
We study the transmission of supersymmetry breaking via
gravitational interactions in a five-dimensional brane-world
compactified on $S^1/Z_2$. We assume that chiral matter and gauge
fields are confined at the orbifold fixed points, where
supersymmetry is spontaneously broken by effective brane 
superpotentials. Using an off-shell supergravity multiplet we
integrate out the auxiliary fields and examine the couplings between 
the bulk supergravity fields and boundary matter fields. The corresponding
tree-level shift in the bulk gravitino mass spectrum induces one-loop
radiative masses for the boundary fields. We calculate the boundary
gaugino and scalar masses for arbitrary values of the brane 
superpotentials, and show that the mass spectrum reduces to the
Scherk-Schwarz limit for arbitrarily large values of the brane 
superpotentials.

\end{abstract}

\end{titlepage}
\setcounter{footnote}{0}
\vskip2truecm

\section{Introduction}

An important question in the supersymmetric standard model is how 
supersymmetry is spontaneously broken in the low-energy world. 
This question has been mainly addressed in the context of 
four-dimensional effective theories with limited success, but 
recently the idea of extra dimensions has allowed for new 
possibilities~\cite{anton}--\cite{gp}. 
The extra dimensional framework is particularly interesting because it 
provides a new geometrical perspective in understanding some of the 
problems of conventional theories. In particular, the nonlocal nature
of communicating supersymmetry breaking across the compact bulk from one
boundary to another can soften the divergences of the soft mass 
spectrum~\cite{aq,adpq}.

The Horava-Witten scenario~\cite{hw} provides the prototype model
in which to study the transmission of supersymmetry breaking via an
extra dimension. In this model the transmission of supersymmetry
breaking can become quite involved, and to calculate boundary 
soft masses one requires the bulk/boundary couplings.
However, the essential features can be captured by studying a 
simpler five-dimensional super-Yang-Mills theory coupled to chiral matter
on the boundary~\cite{mirpes}. In this toy model the couplings between
five-dimensional supermultiplets and four-dimensional boundary fields are
obtained by working with off-shell supermultiplets and including the 
auxiliary fields. As noticed in Ref.~\cite{mirpes} the dimensional 
reduction of bulk fields leads to new couplings between bulk and 
boundary fields which are required for consistency. In particular
this allows for the construction of realistic low-energy 
models with bulk gauge fields in flat~\cite{adpq} and 
warped space~\cite{gp}.

In this work we study brane-world supersymmetry breaking in the case
where only gravity propagates in the bulk while the chiral
matter and gauge fields are confined to the four-dimensional boundaries.  
We assume that due to brane dynamics supersymmetry is spontaneously 
broken by effective brane superpotentials. This causes the tree-level 
gravitino mass spectrum to shift by a constant amount depending on the
values of the brane superpotential~\cite{bfz}. At tree-level 
boundary chiral matter and gauge fields are massless but due to 
their gravitational interactions with the bulk gravitinos they will 
receive a supersymmetry breaking mass at one loop. Just like the 
bulk gauge field case, one can use an off-shell formulation to study 
the bulk gravitational case as well. 

Supersymmetric brane-world scenarios from off-shell supergravity have 
been formulated in Refs~\cite{zucker,kugo}. We will predominantly
use the results in Ref.~\cite{zucker} to study an off-shell formulation
of supergravity in the context of supersymmetry breaking
with brane superpotentials. In particular, we will show
that after integrating out the auxiliary  fields of the off-shell 
supergravity multiplet there are new couplings between bulk and
boundary fields. Just like the bulk gauge field case these couplings are
required in order to obtain a consistent supersymmetric limit.

The one-loop mass spectrum will continuously depend on the
brane superpotential parameter, and due to the nonlocal nature
of the supersymmetry breaking the masses will be finite. In fact
we will see that one particular limit of our mass spectrum is the 
familiar Scherk-Schwarz limit~\cite{ss,aq}. Actually depending on 
the size of the extra dimension our one-loop results can
be of order the anomaly-mediated contributions~\cite{rs,glmr} which 
arise from the one-loop rescaling anomalies. Thus, although we do not
discuss this issue in detail, our results could be relevant in 
solving the tachyonic slepton mass problem~\cite{rs}.

The plan of this paper is as follows: In Section 2, after briefly 
reviewing the bulk vector multiplet case we consider the off-shell
supergravity multiplet coupled to boundary fields. In particular we
show that after integrating out auxiliary fields there are
new couplings between boundary gauge fields and bulk supergravity fields.
Supersymmetry breaking is considered in Section 3, where we derive the
unitary matrix responsible for diagonalising the Kaluza-Klein gravitino  
mass spectrum. This is important for determining the couplings
between the boundary and bulk fields. As a further check, the same 
results will also be 
derived more directly using an explicit five-dimensional calculation.
In section 4 we calculate the one-loop gaugino and scalar masses for
arbitrary values of the brane superpotential. We comment on the 
cancellations that are required for consistency and are
satisfied by the new couplings. Again, for completeness we will
present the calculation of the soft mass spectrum using both 
the Kaluza-Klein sum in four dimensions, and the direct five-dimensional
calculation. Finally, our conclusion and comments will
be presented in Section 5.

\section{Off-shell Bulk Supergravity on $S^1/{Z}_2$}

We start from a  pure $N=2$ five-dimensional Poincar\'e supergravity 
\cite{pure5d}, compactified on an orbifold $S^1/{Z}_2$.
Our model will assume that only gravity propagates in the 
bulk whereas chiral matter and gauge fields will be confined to the 4D
boundaries. Thus all supersymmetry breaking effects will be
transmitted by gravity and in particular the gravitino mass spectrum
will shift. In order to study the transmission of supersymmetry
breaking effects between the 4D boundaries it is necessary to work with 
an off-shell formulation of supergravity~\cite{zucker,kugo}. In this way
all bulk-boundary couplings can be  derived.

A similar procedure for gauge fields and hypermultiplets in the 
bulk was considered in Ref.~\cite{mirpes}.
Before launching ourselves into the more involved 
case of supergravity coupled to boundary chiral and vector 
multiplets, we briefly summarize the procedure and results of  
Ref.~\cite{mirpes} for the case of  a $U(1)$ bulk vector multiplet
which is coupled to a chiral matter multiplet on the boundary. 
This will allow us to emphasize some key features which 
will also be present in the supergravity case.

\subsection{5D Yang-Mills multiplet coupled to boundary chiral matter}

The five-dimensional $U(1)$  multiplet with coupling
constant $g_5$ contains a vector field $A^M$, a real scalar field
$\Phi$, and a gaugino $\lambda^i$. The five-dimensional Yang-Mills 
multiplet is then extended to an
off-shell multiplet by adding an $SU(2)$ triplet $X^a$ of real-vauled
auxiliary fields. Here capitalized indices
$M,N$ run over 0,1,2,3,5, lower-case indices $m$ run over 0,1,2,3, and 
$i$, $a$ are internal $SU(2)$ spinor and vector indices,
 with $i = 1,2$, $a= 1,2,3$.

We now compactify the theory on $S^1/Z_2$ and 
 assign even $Z_2$--parity to the fields
\be
\eta^1_L,\,\, A^m,\,\, \lambda_L^1,\,\, X^3\, ,
\ee 
and odd $Z_2$--parity to the fields
\be
\eta^2_L,\,\, A^5,\,\,\Phi,\,\, \lambda_L^2,\,\, X^1,\,\, X^2~,
\ee
where $\eta^i_L$ is the supersymmetry parameter of the $N=1$ supersymmetry
transformations on the boundary at $x_5=0$.
A simple inspection of the supersymmetry transformations reveals that
the fields $A^m$, $\lambda^1_L$, 
and $(X^3 - \partial_5\Phi)$ transform as the vector, gaugino, and 
the auxiliary $D$-field of a 4D $N=1$ vector multiplet \cite{mirpes}.
It is then obvious  how to couple the five-dimensional gauge multiplet to 
a 4D dimensional chiral multiplet $(\phi,\psi_L,F)$ on the boundary.  
One writes the Lagrangian as 
\begin{equation} 
  S =   \int d^5x \left\{ {\cal L}_5 
        + \sum_i \delta(x_5- x_i^\ast){\cal L}_{4i} \right\}\ ,
\label{totalS}\end{equation}
where the sum includes the  walls at $x^\ast_i = 0, \pi R$. The bulk 
Lagrangian ${\cal L}_5$ is  the standard one for a 5D super-Yang-Mills 
multiplet,
and the  boundary Lagrangian has the standard form of a four-dimensional
chiral model built from the chiral multiplet
and with the gauge fields $(A_m,\lambda_L,D)$ replaced by the boundary values
of the bulk fields $(A_m, \lambda^1_L, X^3 - \partial_5 \Phi)$. 

To determine the couplings of the boundary chiral matter to the 
bulk vector multiplet one has to integrate out the auxiliary fields. 
Integrating out the auxiliary field
$X^3$ gives  a boundary Lagrangian of the form
\begin{equation}
   \int d^4x \left[ - \phi^\dagger (\partial_5\Phi) \phi 
        - \frac{1}{2} g_5^2 (\phi^\dagger\phi)^2 \delta(0) \right]~, 
\label{dofz}
\end{equation}
and one finds new  interaction
terms at the boundaries (apart from the usual ones in $N=1$ 4D Yang-Mills
theory coupled to chiral multiplet)
involving the scalar components of the chiral  multiplet
and  the odd field $\Phi$ \cite{mirpes}.

By including the kinetic term of the field $\Phi$, the singular terms
can be rearranged into a perfect square
\begin{eqnarray}
\label{five}
&&-\int d^5x \left[ \frac{1}{2 g_5^2}(\partial_5\Phi)^2 
+\phi^\dagger (\partial_5\Phi) \phi \,\delta(x_5)
        + \frac{1}{2} g_5^2 (\phi^\dagger 
       \phi)^2 \,\delta^2(x_5) \right]\nonumber\\
&=&\frac{-1}{2g_5^2} \int d^5x \left[\partial_5\Phi+g_5^2\phi^\dagger 
       \phi \,\delta(x_5)\right]^2 \ .
\end{eqnarray}
Varying this action with respect to $\Phi$, one finds that the
background expectation value of $\Phi$ is given by
\begin{equation}
    \partial_5\left\langle{\Phi}\right\rangle =
 - g_5^2 \phi^\dagger\phi \left( \delta(x_5) - {1\over 2\pi R}\right) \ .
\label{Phisol}
\end{equation}
Substituting this solution into the Lagrangian (\ref{five}) one finds
that the various singular terms $\delta(0)$ cancel and one is left with
the usual $D$-term interaction
\begin{equation}
  S = -\frac{1}{2}\int d^5x  
  \frac{g_5^2}{4\pi^2 R^2} (\phi^\dagger \phi)^2=-\frac{1}{4}
  \int d^4x \,g_4^2 (\phi^\dagger \phi)^2~, 
\end{equation}
where $g_4^2=(g_5^2/\pi R)$.
To summarize, we have learned that starting from an off-shell formulation
of five-dimensional 
Yang-Mills theory compactified on $S^1/Z_2$ which is coupled to 
chiral matter on the boundaries, new singular interaction
terms appear after integrating out the auxiliary fields. 
The origin of these terms is clear: they are due to the presence 
of the physical propagating odd field $\Phi$ in the effective 
auxiliary $D$-term on the boundary. At the
level of the effective 4D theory, the singular terms disappear after
we substitute in the Lagrangian the solution of the classical 
equation of motion for the odd field $\Phi$.
It is worth emphasizing though  that the singular terms play a 
crucial role at the quantum level since they provide counterterms 
which are necessary in explicit computations
to maintain supersymmetry~\cite{mirpes}. In particular, the role of the 
interaction term proportional to $\delta(0)$ is
to cancel the singular behaviour induced in diagrams where the
$\Phi$-field is exchanged.

\subsection{The supergravity case}

We now want to extend the analysis of the previous subsection
to the case of supergravity. 
The on-shell supergravity multiplet contains the f\"unfbein 
$e_M^{\;\;A}$, the symplectic Majorana 
gravitino $\Psi_M$, and the graviphoton $A_M$. The 
five-dimensional bulk Lagrangian reads~\cite{pure5d}
\bea
\tilde{{\cal L}}_{bulk} 
& = & - {1\over 2} M_5^3 e_5 R_5  - {1\over 4} M_5 e_5 F_{MN} F^{MN}
- {1 \over 6 \sqrt{6}} \epsilon^{MNOPQ} F_{MN} F_{OP} A_Q \nn \\
&   & + i\, M_5  \epsilon^{MNOPQ} \bar{\Psi}_O \gamma_{PQ} D_M \Psi_N - i 
\sqrt{3 \over 2} {1 \over 2} e_5 F_{MN} \bar{\Psi}^M \Psi^N \nn \\
& & + i \sqrt{3 \over 2} {1 \over 4} \epsilon^{MNOPQ} F_{MN} 
\bar{\Psi}_O \gamma_P \Psi_Q + {\rm 4 \!\! - \!\! fermion \;\; terms} \, ,
\label{lbulk}
\eea
where  the five--dimensional coordinates 
are $x^M=(x^m,x^5)$; $M_5$ is the five--dimensional 
Planck mass (we will set it equal to
one from now on unless otherwise stated); 
$e_5 = \det e_M^{\;\;A}$; $R_5$ is the five--dimensional scalar 
curvature; $e_4 = \det e_m^{\;\;a}$, where the latter are the 
components of the f\"unfbein with four-dimensional indices;
finally, $\epsilon^{MNOPQ} = e_5 \cdot e_A^{\;\;M} e_B^{\;\;N} 
e_C^{\;\;O} e_D^{\;\;P} e_E^{\;\;Q} \epsilon^{ABCDE}$, 
$\epsilon^{mnop} = e_4 \cdot e_a^{\;\;m} e_b^{\;\;n} 
e_c^{\;\;o} e_d^{\;\;p} \epsilon^{abcd}$, $\epsilon^{01235} = 
\epsilon^{0123} = + 1$.

The smallest five-dimensional off-shell supermultiplet contains
48 bosonic and 48 fermionic degrees of freedom. This decomposes
into a minimal multiplet with (40+40) components, containing the 
f\"unfbein, gravitino, graviphoton and several auxiliary fields which 
include an isotriplet scalar $\vec t$, 
an antisymmetric tensor $v_{AB}$, a gauge field ${\vec V}_M$, a spinor 
$\lambda$, and a scalar $C$. In addition one has to introduce a 
compensator multiplet with (8+8) degrees of freedom~\cite{zucker}. 
The compensator
multiplet allows for the breaking of the gauged $SU(2)_R$ symmetry.
There exist several possibilities for the compensator multiplet and we will
assume that it is given by the tensor multiplet containing an isotriplet 
scalar $\vec{Y}$, a spinor $\rho$, a scalar $N$ and a 
vector $W_A$ (which can be expressed as a supercovariant field strength
of a three-form $B_{MNP}$). 
The $SU(2)_R$ symmetry is gauged by the auxiliary field ${\vec V}_M$, and
the gauge is fixed by requiring
\be
   \vec Y= e^u(0,1,0)^T~,
\ee
where $u$ is a scalar field. Thus after gauge fixing the field content 
of the five-dimensional theory is given by
\be
(e_M^A, \Psi_M, A_M, \vec{t},v_{AB}, {\vec V}_M,\lambda,C,u,\rho,N,B_{MNP})~.
\ee
Note that in flat space the choice of the tensor multiplet for 
the remaining (8+8)
components is not unique. However, it is advantageous to use the tensor 
multiplet because it allows  a straightforward generalisation to warped 
spaces. The off-shell bulk Lagrangian is given by~\cite{zucker}
\bea
\label{bulk}
{\cal L}_{bulk} &= & 
e^u\left[-\frac{1}{4}R(\widehat{\omega})_{AB}{}^{AB}+4C-\frac{1}{6}
\widehat{F}_{AB}\widehat{F}^{AB}+v_{AB}v^{AB}+20 
\vec{t}\vec{t}-36{(t^2)}^2 \right.\nonumber\\
&-&\frac{1}{4}\partial^Au\partial_Au - \frac{1}{4}V_A^1V^{A1} - 
\frac{1}{4}V_A^3V^{A3}+8\sqrt{3}\Lambda_5 t^2-
\frac{i}{2}\bar{\Psi}_P\gamma^{PMN}{\cal 
D}_M\Psi_N \nonumber\\
&-& \left. 2i\bar{\Psi}_A\gamma^A\lambda-
\frac{\sqrt{3}\Lambda_5}{4}\bar{\Psi}_A\tau^2
\gamma^{AB}\Psi_B 
- \frac{i}{2}\bar{\Psi}_A\Psi_Bv^{AB}\right] 
-12Nt^2+\sqrt{3} \Lambda_5 N\nonumber\\
& -&\frac{1}{\sqrt{3}}F_{AB}v^{AB} 
-\frac{1}{6\sqrt{3}}\varepsilon^{ABCDE} A_AF_{BC}F_{DE}- 
4\bar{\lambda}\tau^2\rho-2i\bar{\lambda}\gamma^A\Psi_A\nonumber\\
& + &
\frac{1}{2}\bar{\rho}\tau^2\Psi_A\partial^Au-\frac{1}{2}\bar{\rho}\tau^1
\Psi^MV_M^3+\frac{1}{2}\bar{\rho}\tau^3\Psi^MV_M^1+\frac{1}{2}\bar{\Psi}_A
\tau^2\gamma^{AB}{\cal D}_B\rho\nonumber\\
& +& 2i\bar{\rho}\gamma^A\Psi_A t^2 - 2\bar{\Psi}_A\tau^1\gamma^A\rho 
t^3 + 2\bar{\Psi}_A\tau^3\gamma^A\rho 
t^1-\frac{1}{2}\bar{\Psi}_A\tau^2\gamma^{AB}\rho \partial_B u\nonumber\\
& - &\frac{1}{12}\varepsilon^{MNPQR}(V_M^2-2 \Lambda_5 A_M)
\partial_NB_{PQR}-32 \vec{t}\vec{t}-\frac{\sqrt{3}i\Lambda_5 }{2}
\bar{\Psi}_A\gamma^A\rho \nonumber\\
& - &\frac{1}{2}\bar{\rho}\tau^2\gamma^{MN}{\cal D}_M\Psi_N + 
\bar{\rho}\tau^2\gamma_B\Psi_Av^{AB} - \frac{1}{2\sqrt{3}}\bar{\rho}
\tau^2\gamma^{ABC}\Psi_A\widehat{F}_{BC}\nonumber\\
& - &\frac{i}{4\sqrt{3}}\bar{\Psi}_A\gamma^{ABCD}\Psi_B(e^u\widehat{F}_{CD} 
+ \frac{1}{2}F_{CD}) + 
(1-e^u)\bar{\Psi}_A\vec{\tau}\gamma^{AB}\Psi_B\vec{t}\nonumber\\
& +&e^{-u}\left[ - 
\frac{i}{4}\bar{\rho}\gamma^{AB}\rho(v_{AB}+\frac{1}{\sqrt{3}}
\widehat{F}_{AB}) - 3\bar{\rho}\tau^2\rho t^2+W_AW^A-N^2\right.\nonumber\\
& -&\left. \frac{i}{2}\bar{\rho}\gamma^A{\cal D}_A\rho+i\bar{\rho}\Psi_A 
W^A\right]-4C -\frac{1}{2}e^{-2u}(\bar{\rho}\tau^2\rho N 
-\bar{\rho}\tau^2\gamma^A\rho W_A) +{\cal L}_{4F}\, ,
\eea
where $\widehat{F}_{AB}=F_{AB}+i(\sqrt{3}/2)\bar{\Psi}_A\Psi_B$ and 
${\cal L}_{4F}$ contains four-fermion interaction terms. We have
also suppressed all the $SU(2)_R$ indices. All the definitions
of the covariant derivatives can be found in Ref.~\cite{zucker}, and we have 
allowed -- for generality --  the presence of a bulk cosmological 
constant $\Lambda_5$ (which will be set to zero later). Note that it
will turn out that on-shell we have $u=0$, 
as can be seen from the variation of (\ref{bulk}) with respect to $C$.

The fifth dimension is compactified on the orbifold 
$S^1/{Z}_2$, obtained by the identification $x_5 
\leftrightarrow - x_5$. Under the orbifold symmetry fields can
be classified as either even $(P=+1)$ or odd $(P=-1)$.
Distinguishing between four--dimensional and extra--dimensional indices, 
and decomposing the generic five--dimensional spinor $\Psi$ and its
conjugate $\bar{\Psi}$ into four--dimensional ones, with the convention
that $\Psi \equiv (\psi^1_\alpha \, , \, \bar{\psi^2}^{\ad})^T$ and 
$\bar{\Psi} \equiv ( \psi^{2 \, \alpha} \, , \, \bar{\psi^1}_{\ad} )$,
we assign even ${Z}_2$--parity to
\be
\eta^1 \, ,
\;\;\;
e_m^{\;\;a} \, ,
\;\;\;
e_{55} \, ,
\;\;\;
A_5 \, ,
\;\;\;
\psi_m^1 \, ,
\;\;\;
\psi_5^2 \, ,
\;\;\;
v_{a5} \, ,
\;\;\;
\lambda \, ,
\;\;\;
C \, ,
\;\;\;
V_m^3 \, ,
\;\;\;
V_5^1 \, ,
\;\;\;
V_5^2 \, ,
\;\;\;
t^1 \, ,
\;\;\;
t^2 \, ,
\ee
and odd ${Z}_2$--parity to
\be
\eta^2 
\, , \;\;\;
e_5^{\;\;a}
\, , \;\;\;
e_{m 5} 
\, , \;\;\;
A_m
\, , \;\;\;
\psi_m^2
\, , \;\;\;
\psi_5^1
\, , \;\;\;
V_m^1
\, , \;\;\;
V_m^2
\, , \;\;\;
V_5^3
\, , \;\;\;
v_{ab}
\, , \;\;\;
t^3 \, ,
\ee 
where $\eta$ is the supersymmetry parameter (and from now on we will set 
$e_{55}$ to unity unless otherwise stated).

At the ${Z}_2$-fixed points half of the degrees of freedom are
eliminated, reducing the number of supercharges by one 
half. Thus, we can locate three-branes with $N=1$ supersymmetric 
chiral matter content at the orbifold fixed points.
Note that the orbifold also breaks the $SU(2)_R$ symmetry
at the fixed points to a residual $U(1)_R$.
Since in the following we will be interested in the supersymmetric
couplings between the gravitational sector and the boundary matter fields,
we notice that only the fields which are 
even under  the ${Z}_2$-symmetry will
possess such couplings. For instance, only the Kaluza-Klein tower of
$\psi^1_m$ couples to boundary chiral matter.

\subsection{The intermediate multiplet}

The crucial point in understanding how to couple chiral
and vector multiplets to gravity at the boundaries 
is to construct a four-dimensional gravitational multiplet
involving the even fields. One can show that
all the even fields of the $N=2$ 
minimal multiplet form a non-minimal $N=1$ supergravity 
multiplet in four dimensions with $(16+16)$ components~\cite{zucker}. 
If we conveniently define the four-component 4D Majorana spinors as
\be
\psi_m=
\left(\begin{array}{c}
\psi^1_m \\
\bar{\psi}^1_m\end{array}
\right)\,\,\, {\rm and} \,\,\,
\psi_5=
\left(\begin{array}{c}
\psi_5^2 \\
\bar{\psi}_5^2\end{array}
\right)\, ,
\ee
then the $N=1$ supergravity multiplet in four dimensions with 
$(16+16)$ components is given by 
\be
\label{interm}
 (e_m^a,\psi_m,b_a,a_m,\lambda,S,t^1,t^2)~.
\ee
This multiplet (\ref{interm}) is known as the intermediate multiplet and was 
first studied in Ref.~\cite{sw}. 
The auxiliary fields are identified as~\cite{zucker}
\bea
b_a & = &  v_{a 5}~,\\
\label{auxdef}
a_m & = & -\frac{1}{2}( V_m^3 - \frac{2}{\sqrt{3}} 
\widehat{F}_{m5}e^5_{5}+4e_m^av_{a 5})~,\\
S & = & C-\frac{1}{2}e_{5}^5(\partial_5 
t^3-\bar{\lambda}\tau^3\psi_5+V_5^1 t^2-V_5^2t^1)~,
\eea
where in particular the auxiliary field $a_m$ is a combination of 
$V_m^3$, $v_{a5}$ and the (even) field strength 
$\widehat{F}_{m5}=F_{m5}+i\sqrt{3}/2 \bar\psi_m\psi_5$ of the 
propagating graviphoton field, $A_M$.

The situation is then completely analogous to what happens in the 
case of an off-shell bulk vector multiplet in 5D analyzed previously.
There, the presence of the propagating odd field $\Phi$ in the 
effective $D$-term  $D=(X^3-\partial_5 \Phi)$ on the boundary induced 
new interactions between the the chiral matter fields living at the 
boundary and the $\Phi$ field. This suggests that after
integrating out the supergravity auxiliary fields, 
one should expect new interaction
terms at the boundaries (compared to the usual ones in $N=1$ 4D supergravity
coupled to chiral or vector multiplets)
involving the components of the chiral and vector multiplets
and  the field strength $F_{m5}$. A similar argument can be 
made for the even part of the gravitino field $\psi_5$, even though we
anticipate that $\psi_5$  will be set to zero
in the unitary gauge after supersymmetry breaking.

\subsection{Coupling to boundary matter fields}

It is now quite clear how to couple boundary matter fields to
the five-dimensional $N=2$ gravity multiplet. We write
\be
   S=\int d^5 x\left[ {\cal L}_{bulk} +\sum_i \delta(x_5-x^\ast_i)
     {\cal L}_{4i} \right]~,
\ee
where ${\cal L}_{bulk}$ is given by (\ref{bulk}) and the orbifold
fixed points are located at $x^\ast =0$ and $x^\ast = \pi R$. To obtain
the explicit form of the boundary Lagrangian, ${\cal L}_4$ one simply
uses the expressions given in Ref.~\cite{sw} and replaces the auxiliary
fields of the intermediate multiplet with the boundary values given 
by the expressions in Eq. (\ref{auxdef}). 

In particular for an $N=1$ vector multiplet 
$(u_m,\chi,D)$ on the boundary one finds that~\cite{zucker}
\bea
\label{l4gauge}
{\cal L}_4 &=& {\rm Tr}\left[\frac{1}{4} D D
-\frac{1}{4}\widehat{u}_{ab}\widehat{u}^{ab}
+ i\bar{\chi}\gamma^a\widehat{\cal D}_a\chi
-3i\bar{\chi}\gamma^a\gamma^{5}\chi b_a 
-\frac{i}{2}\bar{\psi}_m\gamma^m \gamma^{5}\chi D \right.\nonumber\\
&-& \frac{i}{4}\bar{\psi}_m\gamma^m\gamma^{ab}\chi\widehat{u}_{ab}
+ \left. \frac{1}{4}\bar{\psi}_m\tau^2\gamma^{mn}\psi_n\bar{\chi}\tau^2\chi
+\frac{1}{4}\bar{\psi}_m\tau^2\gamma^{mn}\gamma^{5}\psi_n\bar{\chi}
\tau^2\gamma^{5}\chi\right]~,
\eea
where 
\begin{eqnarray}
\widehat{u}_{ab}&=&u_{ab}-i\bar{\psi}_a\gamma_b\chi+i
\bar{\psi}_b\gamma_a\chi\, ,\nonumber\\
\widehat{\cal D}_m\chi&=&{\cal D}_m\chi-\gamma^5\chi a_m   \, ,
\end{eqnarray}
and ${\cal D}_m\chi$ is the usual covariant derivative for a gaugino
field coupled to gravity.

Following the procedure of Ref.~\cite{mirpes} and setting 
\be
v_{AB}^\prime=v_{AB}-
\frac{1}{2\sqrt{3}}\widehat{F}_{AB}~,
\ee
to canonically normalize the kinetic term of the
graviphoton field strength, we can 
integrate out the auxiliary fields for the case of a boundary vector 
multiplet. The Lagrangian for the auxiliary fields
$V_m^3$ and $v_{a5}^\prime\equiv b_a^\prime$ is given by
\be
{\cal L}_{{\rm aux}} = -2 (b_m^\prime)^2-\frac{1}{4}(V_m^3)^2+
\frac{1}{2} F_{m5}^2
-i \bar\chi\gamma^m\gamma^5\chi\left(b_m^\prime+
\frac{\sqrt{3}}{2} F_{m5}-\frac{1}{2} V_{m}^3\right)\delta(x_5)~,
\ee
which leads to the following vacuum expectation values 
\begin{eqnarray}
V_m^3&=&   i\bar\chi\gamma_m\gamma^5\chi\,\delta(x_5)~, \nonumber\\
b_m^\prime &=& -\frac{i}{4}\bar\chi\gamma_m\gamma^5\chi\,\delta(x_5)~.
\label{con}
\end{eqnarray}
When inserted back into the Lagrangian (\ref{l4gauge}), the 
conditions (\ref{con}) give the usual four-dimensional interaction terms 
of a vector multiplet coupled to gravity plus -- as argued above --
new interaction terms involving the graviphoton field strength, namely
\be
   \int d^5 x \frac{1}{2}\,
  \left( F_{a5} +i \frac{\sqrt{3}}{2}\bar\psi_a\psi_5 
-i\frac{\sqrt{3}}{2}\delta(x_5)\bar\chi\gamma_a\gamma_5\chi\right)^2~.
\ee
After performing an $x_5$ integration, we 
see that singular terms proportional to $\delta(0)$ appear
in the Lagrangian\footnote{The same result can be recovered by a full
on-shell procedure~\cite{lalak}.}, and as expected they contain the 
physical fields $F_{m5}$ and $\psi_5$ which appear in the boundary
auxiliary fields (\ref{auxdef}). This situation is therefore analogous 
to that described in Ref.~\cite{mirpes} where a 5D vector multiplet 
is coupled to chiral matter on the boundary, or to that occurring
in $E_8\times E_8$ strongly coupled heterotic string theory~\cite{hw}.

Since the  singular terms can be
written as a perfect square, at least formally 
they  can be eliminated by a field redefinition 
${\cal F}_{a5}=\widehat{F}_{a5}- i \frac{\sqrt{3} }{2}
\delta(x_5) \bar\chi\gamma_a\gamma^5\chi$. At the
level of the effective 4D theory, the singular terms disappear after
we substitute in the Lagrangian the solution of the classical 
equation of motion for the odd field $A_a$.
However,  the singular terms at the quantum level play a crucial 
role since they provide counterterms which are necessary in explicit 
computations to maintain supersymmetry. 

A similar, but much more involved procedure can also be followed 
to obtain the four-dimensional Lagrangian for a boundary chiral multiplet
$(\varphi,\psi_\varphi,F)$.

\subsection{The tensor multiplet at the boundaries}

We have derived the couplings between the gravity fields
of the intermediate multiplet and the matter fields. There will also be 
couplings between the gravity fields of the intermediate multiplet
and the fields of the tensor multiplet, since in addition to the minimal 
multiplet, parities can also be assigned to the tensor 
multiplet~\cite{zucker}. We assign even ${Z}_2$--parity to
\be
Y^1 \, ,
\;\;\;
Y^2 \, ,
\;\;\;
\rho \, ,
\;\;\;
N \, ,
\;\;\;
B_{mnp} \, ,
\ee
and odd ${Z}_2$--parity to
\be
Y^3 \, ,
\;\;\;
B_{mn5} \,.
\ee
On the boundary the even fields of the tensor multiplet form a chiral
multiplet 
\be
(A,B,\psi,F,G)
\ee 
with chiral weight $w=2$ at the fixed points.
The precise correspondence on the boundary is
\be
\big(A,~B,~\psi,~F,~G\big)= \big(~Y^2,~Y^1,~\rho,~-2N+\widehat{\cal 
D}_{5}Y^3, ~+2 W^{5}+12(t^1Y^2-Y^1t^2)\big)~.
\ee
Using the result for the $F$-term density of 
a chiral multiplet~\cite{sw} this leads to the complete off-shell action
\be
S=\int d^5x~e_5\left\{ {\cal L}_{bulk} 
+ \frac{1}{M_5^3}\left[ W_0 \delta(x_5) + W_\pi\delta(x_5-\pi R)\right]
{\cal L}_{4T}\right\}~,
\label{compaction}
\ee
where $W_0$ and $W_\pi$ are complex constants with dimension of (mass)$^3$, 
and the boundary Lagrangian is given by
\be
{\cal L}_{4T} =-2N + e^u  V_5^1 - \bar{\rho}\tau^3\psi_5
+i\bar{\psi}_m\gamma^m\rho+\frac{1}{2}e^u 
\bar{\psi}_a\tau^2\gamma^{ab}\psi_b - 12 e^u t^2~.
\ee
Eliminating  the auxiliary fields in (\ref{compaction}) we 
finally obtain the on-shell action
\be
S =\int d^5 x \left\{ \widetilde{\cal L}_{bulk} +\frac{1}{2 M_5^3}
\left[W_0 \delta(x_5)+ W_\pi \delta(x_5-\pi R)\right]
\bar{\psi}_a\tau^2\gamma^{ab}\psi_b\right\}~,
\ee
where we have set $\Lambda_5=0$ and 
${\widetilde{\cal L}}_{bulk}$ is the on-shell Lagrangian of bulk 
supergravity (\ref{lbulk}). The Killing spinor equation
$\delta\psi_M=0$ reduces to 
\be
\label{killing}
\partial_5\eta=-i\left[W_0 \delta(x_5)+ W_\pi \delta(x_5-\pi R)\right]
\gamma^5 \tau^2\eta~.
\ee
Decomposing the 5D symplectic spinor $\eta$ into two component objects
$\eta_i^T=(\eta^+_i,\eta^-_i)$ ($i=1,2$), Eq. (\ref{killing}) 
 has a non trivial solution only if $W_0+W_\pi=0$. 
The solution is: $\eta^+_1=\epsilon$ and $\eta^-_2=-i\theta(x^5)
\epsilon$, where $\epsilon$ is a four-dimensional Weyl spinor
which generates the supersymmetry of the ground state.
Therefore, the flat space solution is supersymmetric provided
that $W_0+W_\pi=0$ \cite{zucker}.

\section{Supersymmetry breaking }

We  now  consider the case in which supersymmetry is broken. 
If $W_0+W_\pi\neq 0$ then we will see that the flat space solution 
spontaneously breaks supersymmetry. The 
supersymmetry breaking will be transmitted to matter on branes located 
at the orbifold fixed points via gravity. The brane action is assumed to be
\bea
\label{sbrane2}
S_{brane} &=& \frac{1}{2}
\int d^4 x \int_{- \pi R}^{+ \pi R} 
d x_5 \,  e_4 \, \left\{ \delta(x_5) {\cal L}_4^{(0)}
+\delta(x_5-\pi R) {\cal L}_4^{(\pi)} \right.\nonumber\\
&+&\left. \frac{1}{2 M_5^3}\left[ \delta(x_5) 
W_0 + \delta(x_5 - \pi R) W_{\pi}  \right]
\psi_m^1 \sigma^{mn} \psi_n^1 + {\rm \; h.c.}\right\}~,
\eea
where ${\cal L}_4^{(i)}$ are the boundary Lagrangians describing the 
interaction of matter with the bulk.
Expanding the fermions in Fourier modes   consistently with
their boundary conditions and ${Z}_2$--parity assignments leads to
\bea
\psi^+ (x_5) & = & {1 \over \sqrt{\pi R}} \left[ \psi^{+}_{0} +
\sqrt{2} \sum_{\rho=1}^{\infty} \psi^{+}_{\rho} \cos\frac{\rho}{R} x_5
\right] \, , \nn \\
\psi^- (x_5)
& = &
{1 \over \sqrt{\pi R}} \left[
\sqrt{2} \sum_{\rho=1}^{\infty} \psi^{-}_{\rho} \sin\frac{\rho}{R}x_5
\right] \, ,
\label{modez2}
\eea
where $\psi^+$ stands for $(\psi_m^1,\psi_5^2)$ and $\psi^-$
for $(\psi_m^2,\psi_5^1)$. The presence of the brane superpotential
induces a  mixing between the different Kaluza-Klein levels. 
The fields $\psi^1_{5,\rho}$, $\psi^2_{5,0}$ and $\psi^2_{5,\rho}$ 
($\rho>0$) are Goldstinos, eaten up by the gravitinos via the 
super--Higgs effect~\cite{mno,bfz}.
As we saw earlier when $W_0+W_\pi=0$ it is still possible to define a 
Killing spinor and the $N=1$ supersymmetry is not spontaneously broken 
by the presence of the brane superpotentials. This means that the amount of 
supersymmetry breaking is fixed by the order parameter 
$F=(W_0+W_\pi)/M_5$.

The infinite-dimensional gravitino mass matrix can be easily 
diagonalized~\cite{bfz,ddg}. Defining
\be
\psi_{m,\rho}^{\pm} = \frac{1}{\sqrt{2}}(\psi_{m,\rho}^1 \pm 
\psi_{m,\rho}^2) \, , \; (\rho > 0) \, ,
\;\;\;\;\;\;
{\cal P}_{\pm} = {1 \over 2 \pi M_5^3}
\left( W_0 \pm W_\pi \right)
\, ,
\ee
one finds that the modes of $\psi_m^1$ and $\psi_m^2$ combine
to form nearly degenerate pairs of Majorana states~\cite{bfz} with masses
\be
\label{eigval}
{\cal M}_{3/2}^{(\rho)} = {1 \over R}
\left(\rho+\Delta  \right) \, ,
\;\;\;\;\;
(\rho=0,\pm1,\pm2,\ldots) \, ,
\ee
where
\be
\label{del}
\Delta={1 \over \pi}  \arctan \left[
4 \pi \cP_+  \over \pi^2 (\cP_-^2 - \cP_+^2) + 
4 \right]~.
\ee
Note that the mass eigenvalues (\ref{eigval}) are both positive
and negative. Of course, the absolute values give the physical masses,
$\frac{\Delta}{R}$, $\frac{1\pm\Delta}{R}$, $\frac{2\pm \Delta}{R}\dots$,
where the physical range of the supersymmetry breaking parameter $\Delta$
is $0\leq \Delta \leq 1/2$.
Notice that in the supersymmetric limit, $W_0 =- W_\pi$ $(\cP_+=0)$ 
the gravitino mass spectrum remains unshifted as expected.

After the super-Higgs mechanism,
from the four-dimensional point of view, the physical spectrum
contains one massless $N=1$ gravitational multiplet with spins $(2,3/2)$
built up with the zero modes of $e_m^a$ and $\psi^1_m$; one radion multiplet
composed of the zero modes $e_{55}$, $A_5$ and $\psi_5^2$ and an infinite
series of massive multiplets of $N=2$ supergravity with spins
$(2,3/2,3/2,1)$.

Let us now consider two interesting physical limits in the instance
where $W_0$ vanishes identically and $W_\pi$ is nonzero.
This means that $\cP_+=-\cP_-=\frac{W_\pi}{2\pi M_5^3} $.
In this case the only source of supersymmetry breaking 
appears as a constant superpotential on the ``hidden'' brane.

\vskip 0.5cm

{\it i)}~If the absolute value of the 
superpotential $\left|W_\pi\right|$ is much smaller than $M_5^3$, 
$\left|W_\pi\right|\ll M_5^3$, the function $\Delta$ is
well approximated by  $\cP_+=\frac{W_\pi}{2\pi M_5^3}$.
This means that the gravitino zero-mode mass is given by 
\be
\label{grav0}
{\cal M}_{3/2}^{(0)} = \frac{W_\pi}{M_4^2}~,
\ee
where we have invoked the relation $M_4^2\simeq M_5^3 \pi R$. This
is the familiar four-dimensional expression for $N=1$ supergravity.
The other massive modes are well separated from the lowest mode by 
a multiple of $R^{-1}$. This means that the low-energy 
effective four-dimensional theory (describing the
physics below the scale $R^{-1}$ and consisting only of zero modes) 
is simply the $N=1$ supergravity theory with supersymmetry spontaneously
broken by the nonvanishing superpotential $W_\pi$.

\vskip 0.5cm

{\it ii)}~If the absolute value of the superpotential $\left|W_\pi\right|$ is 
much larger than $M_5^3$, $\left|W_\pi\right|\gg M_5^3$), then the function 
$\Delta$ is well approximated by  $1/2$. This means that the gravitino
zero-mode mass is given by 
\be
\label{gravinf}
{\cal M}_{3/2}^{(0)}=\frac{1}{2R}.
\ee
Notice that the zero-mode gravitino mass no longer depends 
on the superpotential 
parameter. All the massive modes are again separated from the 
zero mode by a multiple of $R^{-1}$. 
Therefore, the gravitino mass spectrum is identical to that obtained
from the Scherk-Schwarz supersymmetry breaking mechanism~\cite{ss} 
which makes use of non-trivial (anti-periodic) boundary conditions
for the five-dimensional gravitino field upon compactification of the
fifth dimension. This observation will turn out to be useful 
in the following  when we will show how   
supersymmetry breaking is communicated to the visible brane 
at one loop through the gravitational sector living in the bulk. 
Indeed, mass splittings for the observable fields
have been computed in the context of M-theory~\cite{aq}
where  supersymmetry breaking by gaugino condensation in the 
strongly coupled heterotic string can be described by an analogue of
Scherk-Schwarz compactification on the eleventh dimension. 
At the lowest order, supersymmetry is broken only in the 
gravitational and moduli sector 
at a scale $m_{3/2}\sim R^{-1}$, where $R$ is the radius of the 
eleventh dimension, and it is transmitted to the observable world by 
 gravitational interactions. We will therefore
be able to reproduce the results 
of Ref.~\cite{aq} in the limit $\left|W_\pi\right|\gg M_5^3$.

\subsection{From the interaction to the mass gravitino eigenstates}

The infinite unitary matrix $\U$ which diagonalizes the infinite 
gravitino mass matrix, ${\cal M}_{3/2}$ is defined by 
\be
\U\,{\cal M}_{3/2}\, \U^\dagger={\cal M}_D,
\ee
where ${\cal M}_D$ is the diagonal mass matrix whose eigenvalues 
are given in (\ref{eigval}). 
Knowledge of the unitary matrix $\U$ is necessary in order
to perform  the one-loop computation of the soft supersymmetry
breaking masses of the fields living on the boundaries because the
interactions are not in the mass eigenstate basis.
Correspondingly, the gravitino mass 
eigenstates $\widetilde{\psi}$ are obtained from the relation
\be
\widetilde{\psi}=\U\,\psi \, ,
\ee
where $\psi=(\psi_0^1,\psi_1^+,\psi_1^-,\psi_2^+,\psi_2^-,\cdots)$ 
represents the infinite gravitino eigenvector for the mass matrix 
${\cal M}_{3/2}$. For arbitrary values of the brane superpotentials, 
$W_0$ and $W_\pi$, the gravitino mass eigenvector $\widetilde{\psi}_\rho$ 
with mass eigenvalue ${\cal M}_{3/2}^{(\rho)}$ is given by
\be
\widetilde{\psi}_\rho= N_{\lambda^{(\rho)}}\,\left(1,
\frac{\lambda^{(\rho)}\xi}{\lambda^{(\rho)}-1},
\frac{\lambda^{(\rho)}\xi}{\lambda^{(\rho)}+1},
\frac{\lambda^{(\rho)}}{\lambda^{(\rho)}-2},
\frac{\lambda^{(\rho)}}{\lambda^{(\rho)}+2},
\frac{\lambda^{(\rho)}\xi}{\lambda^{(\rho)}-3},
\frac{\lambda^{(\rho)}\xi}{\lambda^{(\rho)}+3},
\cdots\right)\, ,
\ee
where $\lambda^{(\rho)}=\rho+\Delta$, $N_{\lambda^{(\rho)}}$ is a 
normalisation constant and 
\be
  \xi=\frac{2\cP_+ \cP_-}{\cP_+^2 + \cP_-^2+(\cP_+^2- \cP_-^2)
      (1+\pi \cP_+ \tan\frac{\Delta\pi}{2})}~.
\ee
Requiring that the vector $\widetilde{\psi}_\rho$ has norm equal to unity
gives for the normalisation constant
\be
 N_{\lambda^{(\rho)}}=\frac{1}{\lambda^{(\rho)}\pi}
\frac{\sqrt{2} \sin\lambda^{(\rho)}\pi}{\sqrt{1+\xi^2 +(1-\xi^2)
\cos\lambda^{(\rho)}\pi}}\, .
\ee
The matrix $\U$ can therefore be written as
\be
\U=\left[
\begin{array}{ccccc}
N_{\lambda^{(0)}} & \frac{\lambda^{(0)} N_{\lambda^{(0)}}\xi} 
{\lambda^{(0)}-1} & \frac{\lambda^{(0)} N_{\lambda^{(0)}}\xi}
{\lambda^{(0)}+1} &  \frac{\lambda^{(0)} N_{\lambda^{(0)}}}
{\lambda^{(0)}-2} & \ldots\\
N_{\lambda^{(1)}} & \frac{\lambda^{(1)} N_{\lambda^{(1)}}\xi}
{\lambda^{(1)}-1} & \frac{\lambda^{(1)} N_{\lambda^{(1)}}\xi}
{\lambda^{(1)}+1} & \frac{\lambda^{(1)} N_{\lambda^{(1)}}}
{\lambda^{(1)}-2} & \ldots\\
N_{\lambda^{(2)}} & \frac{\lambda^{(2)} N_{\lambda^{(2)}}\xi}
{\lambda^{(2)}-1} & \frac{\lambda^{(2)} N_{\lambda^{(2)}}\xi}
{\lambda^{(2)}+1} & \frac{\lambda^{(2)} N_{\lambda^{(2)}}}
{\lambda^{(2)}-2} & \ldots\\
N_{\lambda^{(3)}} & \frac{\lambda^{(3)} N_{\lambda^{(3)}}\xi}
{\lambda^{(3)}-1} & \frac{\lambda^{(3)} N_{\lambda^{(3)}}\xi}
{\lambda^{(3)}+1} & \frac{\lambda^{(3)} N_{\lambda^{(3)}}}
{\lambda^{(3)}-2} & \ldots\\
\vdots & \vdots & \vdots & \vdots &\ddots\\
\end{array}\right]~.
\ee
It is not difficult to check that this matrix is unitary. 
In addition an infinite sum over the unitary matrix elements can 
be performed and leads to the result
\be
\sum_{n=-\infty}^{\infty}\, (\pm 1)^n \U_{kn} =
\frac{1\mp\xi+(1\pm\xi)\cos\lambda^{(k)}\pi}
{\sqrt{2}\sqrt{1+\xi^2+(1-\xi^2)\cos\lambda^{(k)}\pi}}~,
\ee
and similarly for $\sum_{n}\, (\pm 1)^n \U^{*}_{kn}$.
In particular for $W_0=0$ and $W_\pi \neq 0$ (i.e. $\xi =-1$)
we obtain
\be
\label{prop}
\sum_{n=-\infty}^{\infty}\, \U_{kn}=1 \qquad {\rm and} \qquad
\sum_{n=-\infty}^{\infty}\, (-1)^n \U_{kn}=\cos\lambda^{(k)}\pi~,
\ee
while for $W_0\neq 0$ and $W_\pi=0$ (i.e. $\xi =1$) we have
\be
\label{prop1}
\sum_{n=-\infty}^{\infty}\, \U_{kn}=\cos\lambda^{(k)}\pi
\qquad {\rm and} \qquad
\sum_{n=-\infty}^{\infty}\, (-1)^n \U_{kn}=1~.
\ee
These relations will be useful later when we consider the
gravitational interaction of gravitinos with boundary matter.

\subsection{5D interpretation}

Before closing this section, we wish to give an alternative 
derivation of the mass eigenvalue formula (\ref{eigval}), and a more
transparent explanation of the relations (\ref{prop}) and (\ref{prop1}). 
Therefore, let us consider the equations of motion of the five-dimensional
gravitino fields after we have set $W_0=0$ and chosen the unitary gauge 
(the opposite case in which $W_\pi=0$ can be analyzed in a similar fashion).
The gravitino equations of motion read
\begin{eqnarray}
\label{one}
\epsilon^{mnpq}\bar{\sigma}_n\partial_p\psi^1_q+
2\,\bar{\sigma}^{mn}\partial_5\bar{\psi}^2_n+
2\,\frac{W_\pi}{M_5^3}\,\delta(x_5-\pi R)\bar{\sigma}^{mn}
\bar{\psi}^1_n&=&0~,\\
\label{two}
\epsilon^{mnpq}\bar{\sigma}_n\partial_p\psi^2_q-
2\,\bar{\sigma}^{mn}\partial_5\bar{\psi}^1_n&=&0\, .
\end{eqnarray}
If we now write $\psi_n^i(x_a,x_5)=\eta_n^i(x_a)f_i(x_5)$ ($i=1,2$), 
and integrate Eq. (\ref{one}) in the interval 
$(\pi R-\varepsilon,\pi R+\varepsilon)$ we obtain
\be
\label{find}
\eta_n^2(x_a)=\frac{W_\pi}{2 M_5^3}\frac{f_1(\pi R)}
{f_2(\pi R)}\,\eta_n^1(x_a)\, .
\ee
The $\eta_n^i$ component of the gravitino satisfies 
the Rarita-Schwinger equation
\be
\epsilon^{mnpq}\bar{\sigma}_n\partial_p\eta^i_q=2\, \frac{\lambda}{R}
\bar{\sigma}^{mn}\bar{\eta}^i_n~,
\ee
and using the relation (\ref{find}) one gets from Eqs. (\ref{one}) and
(\ref{two}) that $f_1(x_5)\propto \cos\left(\frac{\lambda}{R} x_5\right)$
and $f_2(x_5)\propto \sin\left(\frac{\lambda}{R} x_5\right)$
together with the consistency relation
\be
\label{tanrel}
\tan(\lambda \pi) =\frac{W_\pi}{2 M_5^3}~.
\ee
The solution of this equation reproduces the Kaluza-Klein mass spectrum 
$\lambda^{(\rho)}=\rho+\Delta$ given in Eq. (\ref{eigval}) for
${\cal P}_+=-{\cal P}_-=\frac{W_\pi}{2\pi M_5^3}$. Note also 
that (\ref{find}) reduces to $\eta_n^2(x_a)=\eta_n^1(x_a)$ when
we use the relation (\ref{tanrel}).

From this 5D picture we can also extract a transparent interpretation
of the  relations (\ref{prop}). Since only the even gravitino $\psi^1_m$
couples to the boundaries, the interaction between the 
gravitino mass eigenstates $\eta_m^1(x_a)$ and the boundary chiral 
and vector multiplets are accompanied by the five-dimensional
wave-function $f_1(x_5)\propto \cos\left(\frac{\lambda}{R} x_5\right)$. Such
a wave-function is equal to unity if matter and vector multiplets
live on the $x_5=0$ boundary, or alternatively 
equal to $\cos(\lambda^{(\rho)}\pi)$ if
they live on the $x_5=\pi R$ boundary. This is precisely equivalent 
to the relation (\ref{prop}).
We can also alternatively describe the limit in which there is a 
large supersymmetry breaking on the brane at $x_5=\pi R$, $W_\pi\gg
M_5^3$. In this case the eigenvalue $\lambda^{(\rho)}$ tends to the 
value $\left(\rho+\frac{1}{2}\right)$ and the wavefunction of the gravitino 
is suppressed at the brane where supersymmetry breaking occurs. In other 
words, for simple energetic reasons, the gravitino prefers to live in the 
bulk as far away as possible from the boundary where it acquires a large mass.

\section{Communication of supersymmetry breaking via the bulk}

As we have seen in the previous section, the introduction of a constant
superpotential $W_\pi$ on the brane located at $x_5=\pi R$ induces a breaking
of supersymmetry in the five-dimensional gravitational sector, while it 
remains unbroken in the visible sector living on the brane located
at $x_5=0$. The communication of supersymmetry breaking to the
visible sector is then expected to arise radiatively via gravitational 
interactions. This is the issue which we will study below.

\subsection{Boundary vector multiplet}

Let us consider a vector supermultiplet 
$(u_m,\chi, D)$ on the boundary at $x_5=0$. Our goal is to study 
how supersymmetry breaking on the brane at $x_5=\pi R$ is transmitted 
by gravity to the boundary vector supermultiplet.

The coupling between the intermediate multiplet representing
bulk gravity and boundary vector multiplets is given in 
(\ref{l4gauge}). At tree-level the gauginos are all massless, but the
interaction with gravitinos will induce a one-loop radiative mass. From
the parity assignments in (\ref{modez2}) only the even gravitino modes,
$\psi_{m,\rho}^1$ will couple to the boundary gauginos.
Inspection of the off-shell Lagrangian does 
not reveal any new couplings between gaugino fields and the gravitino.
Instead the new couplings that do appear, such as the $\delta(0)$ terms,
are required in order to obtain the usual result consistent
with the $N=1$ supersymmetric limit.

The interaction between the gauginos and the gravitons do not
give any contribution to the gaugino mass (as can be easily understood
from chirality arguments). Therefore, there is no 
supersymmetric cancellation between graviton and gravitino
loops; the  gravitino contributions
have to sum up and give a finite result~\cite{barbieri}.

\begin{figure}[htb]
\begin{center}
\epsfxsize=4in\leavevmode\epsfbox{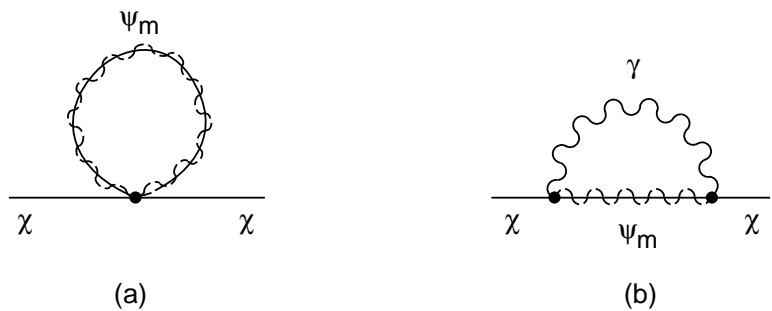}
\end{center}
\caption{\it The one-loop diagrams containing the gravitino $\psi_m$
which give contributions to the gaugino mass.}
\label{gauginofig}
\end{figure} 

We work in the harmonic gauge $\gamma^m\psi_m=0$ 
(integrating out the auxiliary field $\lambda$ in the bulk Lagrangian 
(\ref{bulk}) gives $\gamma^A\psi_A=0$ which reduces to 
$\gamma^m\psi_m=0$ for the gauge choice $\psi_5=0$. Note that
there is no $\lambda$ dependence in ${\cal L}_{4F}$~\cite{zucker}).
In $N=4+\epsilon$ dimensions, every Kaluza-Klein state
of the  gravitino field  gives for the gravitino and gauge loops
of Fig.~\ref{gauginofig}, the respective contributions
\begin{eqnarray}
\label{bb}
i\delta \Sigma_{1a}&=&\frac{(3+\epsilon)}{12}
\,\int\,\frac{d^N k}{(2\pi)^N}\,\left[m_n(4+\epsilon)(3+
\epsilon)-2\,\frac{k^2}{m_n}(2+\epsilon)\right]
\frac{1}{k^2+m_n^2}\, ,\\
i\delta \Sigma_{1b}&=&\frac{(3+\epsilon)}{12}
\,\int\,\frac{d^N k}{(2\pi)^N}\,\left[-m_n(12+3\epsilon
-\epsilon^2)+2\,\frac{k^2}{m_n}(2+\epsilon)\right]
\frac{1}{k^2+m_n^2}\, .
\label{bbb}
\end{eqnarray}
We see that the leading  divergent parts going like $k^2$ cancel exactly.
This means that every Kaluza-Klein gravitino state gives a 
 contribution to the gaugino mass $m_\chi$
of the form
\be 
\label{sum}
\epsilon
\,\int\,\frac{d^N k}{(2\pi)^N}\,
\frac{m_n}{k^2+m_n^2}\, .
\ee
The next step is to isolate the divergent pieces in the 
(sum of the) integrals
(\ref{sum}). To do that, one can simplify the sum over the Kaluza-Klein states
using the contour trick from finite temperature field theory which
allows one to identify the different possible divergences~\cite{nibbelink}. 
We can write the sum as 
\be\label{summ}
\sum_{n=-\infty}^{\infty}\,\frac{m_n}{k^2+m_n^2}=
\frac{1}{2\pi i} \int_{\cal C}\, d k_5\,\frac{k_5}{k^2+k_5^2}\, 
{\cal P}(k_5,\Delta)\, ,
\ee
where the contour $\cal C$ is the line from the left to the right  below the
real axis and another line from the right to the left above this axis
and enclosing the poles at $k_5=(n+\Delta)/R$ of the function~\cite{nibbelink}
\be
{\cal P}(k_5,\Delta)=\frac{1}{2}\,
\frac{\pi R}{\tan\left[\pi 
R\left(k_5-\frac{\Delta}{R}\right)
\right]}\, .
\ee
Notice that in the limit of exact supersymmetry $(\Delta=0)$ the sum 
(\ref{summ}) vanishes identically as it should. 

The expression  (\ref{summ}) can be rewritten as
\be
\frac{1}{2\pi i}\,\int_{-\infty}^{+\infty}dk_5\,\frac{k_5}{k^2+k_5^2}\,
\left[{\cal P}(k_5,\Delta)-
{\cal P}(k_5,-\Delta)\right]\, .
\ee
The dimensionally regulated integrals usually split into two pieces:
a pure 5D divergent piece and a finite piece. The
constant parts $\pm\frac{1}{2}\pi R$ of the pole functions 
${\cal P}(k_5,\pm \Delta)$ 
induce the pure 5D divergent pieces, but -- since they do not depend upon 
$\Delta$ -- they sum up to zero. The remainders $\rho_{\pm\Delta}$ of the 
functions ${\cal P}(k_5,\pm \Delta)$  are highly convergent functions and
each gives a finite contribution to the momentum integration 
and -- therefore -- 
no contribution to the gaugino masses after we set $\epsilon=0$. 
One can understand this
result using the analogy with what happens in 4D \cite{barbieri}. There 
gaugino masses are nonvanishing at one-loop if there is no physical cut-off
in the theory. However, if a physical cut-off $\Lambda_c$ 
is present (such as the scale of
gaugino condensation) one has to cut off the integrals (\ref{bb}) and 
(\ref{bbb}) at $k^2=\Lambda_c^2$ and work in exactly four dimensions
($\epsilon=0$). The two contributions (\ref{bb}) and 
(\ref{bbb}) then exactly cancel and we are left with no contribution
to gaugino masses from massive gravitinos. In 5D
the  five-dimensional  divergent pieces cancel and each  finite 
loop  contribution to gaugino masses
can be seen as potentially 4D  divergent integrals made finite by
setting a cutoff at $p\sim R^{-1}$. 
Therefore we find that the gaugino mass do not get any contribution at
the one-loop order if gravity is the mediator of supersymmetry breaking
through the bulk. This result will be confirmed in subsection 4.3 by
a full 5D computation.

Note that if the gauginos were living on the
3-brane at $x_5=\pi R$ then the diagrams in Fig.~\ref{gauginofig}
would be proportional to the
sum $\sum_{n,m}(-1)^{n+m}\sum_k \,\U^{*}_{nk} \U_{km}
\widetilde{\psi}_k\widetilde{\psi}_k$. In this case the sum would 
reduce to $\sum_{k}\cos^2(\lambda^{(k)}\pi)\widetilde{\psi}_k
\widetilde{\psi}_k$. As we learned at the 
end of section 3, the $\cos^2(\lambda^{(k)}\pi)$
factor represents the 
wave-function suppression of the gravitino at the
boundary where supersymmetry is broken. 
In such a case, the gravitino sum  
is multiplied by the coefficient $\beta_n=\cos^2[(n+\Delta)\pi]$.
Repeating the procedure adopted above, we again find that gaugino masses
vanish at one-loop.

\subsection{Boundary chiral multiplet}

We now add a generic matter chiral supermultiplet 
$(\varphi,\psi_\varphi,F_\varphi)$ on the boundary at $x_5=0$. 
The tree-level scalar masses are zero and will again
be induced at the one-loop level. We assume that the 
supersymmetry breaking is on the brane at $x_5=\pi R$ and is transmitted 
by gravity to the boundary matter supermultiplet. 

The interaction terms 
 between bulk gravity and  the chiral multiplet on the boundary
can be found by coupling the supergravity intermediate multiplet 
(\ref{interm}) to the chiral multiplet. The resulting Lagrangian
is quite involved \cite{sw} and we do not report it here. Integrating
out all the auxiliary fields turns out to be a complicated task. 
Apart from the usual
interaction terms present in the $N=1$ 4D supergravity 
Lagrangian coupled to chiral matter, 
new singular interaction terms appear. 
However, as we already explained in section 2,
these new terms may involve only the field strength $\widehat{F}_{m5}$ and
the gravitino component $\psi_5$.  In particular, 
the field strength and  the  current
$J^m=i(\varphi^\dagger\partial^m\varphi-\varphi\partial^m\varphi^\dagger)+
\psi_\varphi\sigma^m\bar{\psi}_\varphi$ 
combine to form again a perfect square
\be
\label{neww}
   \int d^5 x \frac{1}{2}\,
  \left( F_{a5} +i\frac{\sqrt{3}}{2}\bar\psi_a\psi_5
 - i \sqrt{\frac{3}{2}}\, J_a\,\delta(x_5)\right)^2~.
\ee
These interaction terms induce a one-loop correction to the scalar
masses. In particular, the diagram where 
the odd graviphoton field $A_m$ is exchanged leads to
singular behaviour which is however cancelled by the singular 
$J_a\,J^a \delta(0)$ term. This can be seen by writing $\delta(0)$ as
\be 
\delta(0)=\frac{1}{2\pi R} \sum_{n=-\infty}^{\infty}\frac{k^2-(n/R)^2}
{k^2-(n/R)^2}\, .
\ee
We find 
\begin{eqnarray}
\label{nnn}
\delta m_\varphi^2&\propto&\, \frac{1}{M_5^3}
\sum_{n=-\infty}^{\infty}\, \int\,
\frac{d^4 k}{(2\pi)^4}\,\left(\frac{1}{2\pi R}\frac{(n/R)^2}{k^2-(n/R)^2}+
\delta(0)\right)~,\nonumber\\
&=&
\frac{1}{2 M_4^2}
\sum_{n=-\infty}^{\infty}\, \int\,
\frac{d^4 k}{(2\pi)^4}\,\left(\frac{k^2}{k^2-(n/R)^2}
\right)\, ,
\end{eqnarray}
where we have used the relation $M_4^2 \simeq M_5^3 \pi R$.
This contribution to the scalar masses, when summed up to the 
contribution of the  whole tower of massive gravitino states
belonging to the $N=2$ multiplet in 4D, gives a finite result. Similarly,
the diagram where 
the even  field $A_5$ propagates is cancelled, in the supersymmetric limit,
by the contribution from the $\psi_5^2$ gravitino (they live in the
same radion multiplet).
In the unitary gauge the gravitino component $\psi_5$ is eaten up, 
and cannot induce such a cancellation. Its role is then played by the 
lightest mode of the $\psi_m^1$ gravitino.

After these general comments we now proceed to the explicit
computation of the one-loop scalar masses induced by gravity
after supersymmetry breaking. It is important to notice
that  the coupling between the supergravity intermediate multiplet 
(\ref{interm}) and  the boundary chiral multiplet contains
a generic K\"{a}hler potential $\Omega(\varphi,\varphi^\dagger)$ of 
the scalar fields $\varphi$. Thus, the procedure described in 
Ref.~\cite{sw} gives rise to noncanonical kinetic terms of the form
\begin{eqnarray}
\label{matter}
S_{{\rm kin}}=\frac{1}{2 M_5^3}\int\,d^4x\,\int_{-\pi R}^{\pi R}\,
dx_5 e_4\delta(x_5)\,\left\{
\frac{1}{6} \Omega \left[R_4 -\frac{1}{2}\epsilon^{klmn}
\left(\bar{\psi}^1_k\bar{\sigma}_l\psi^1_{mn}-
\psi^1_k\sigma_l\bar{\psi}^1_{mn}\right)\right]\right.\nonumber\\
-\left.\frac{1}{\sqrt{2}}\left(\frac{\Omega_\varphi}{\Omega}
\psi_\varphi\sigma^{mn}\psi^1_{mn}+{\rm h.c.}\right)\right\}~,
\end{eqnarray}
where 
\begin{eqnarray}
\psi^1_{mn}&=&\partial_m\psi^1_n-\partial_n\psi^1_m~,\\
\Omega_\varphi&=&\frac{\partial\Omega}{\partial\varphi}~,
\end{eqnarray} 
and $R_4$ is the Ricci scalar computed using the vierbein $e^a_m(x_5=0)$ 
of the intermediate multiplet.

At this stage one is free to perform a Weyl rescaling that renders the
gravity and gravitino kinetic terms canonical. However, we prefer to 
compute the one-loop scalar masses in the unrescaled Weyl basis
using the Lagrangian (\ref{matter}). This choice is dictated by the fact that
in the unrescaled Weyl basis both gravitons and gravitinos give rise
to one-loop scalar masses and the supersymmetric cancellations
are more transparent. Furthermore, in this basis there are no
direct couplings between the radion supermultiplet containing the 
even fields $e_{55}$, $A_5$ and $\psi^2_5$ (the radion,
for instance, arises from 
fluctuations of $g_{55}$; by  general covariance it can only couple to 
the $55$-component of the matter energy-momentum tensor which vanishes 
for chiral matter on the brane~\cite{ls}). On the contrary,  
in the  Weyl rescaled basis gravitons do not give a contribution 
to the scalar masses if we start from vanishing tree-level scalar 
masses and supersymmetric cancellations are hidden.

\begin{figure}[htb]
\begin{center}
\epsfxsize=4in\leavevmode\epsfbox{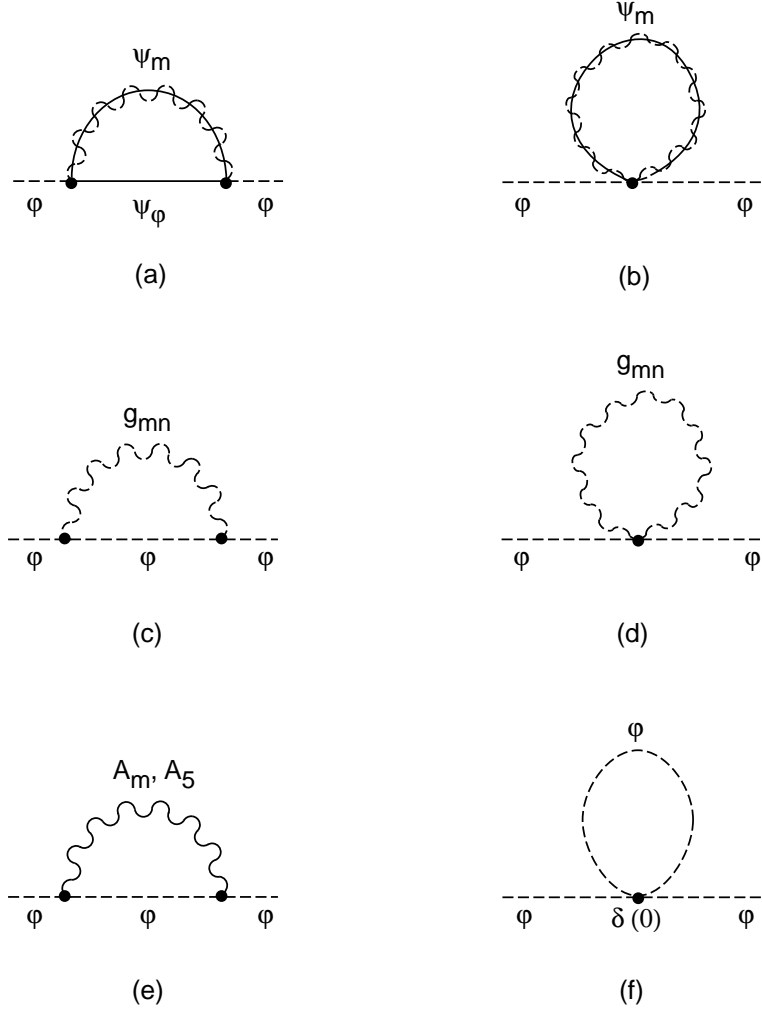}
\end{center}
\caption{\it The one-loop diagrams which give contributions 
to the scalar mass.}
\label{scalarfig}
\end{figure} 

At the one-loop level, the diagrams that contribute to the scalar masses 
are those shown in Fig.~\ref{scalarfig}, 
where the graviton and gravitino vertices come from the kinetic
terms (\ref{matter}). As we already noticed, the fields from the 
boundary in the $N=1$
chiral multiplet $(\varphi,\psi_\varphi, F_\varphi)$, 
always appear in pairs, as dictated by the ${Z}_2$ invariance. 
Just like the gaugino case, only the parity even gravitinos couple
to the boundary chiral multiplet.

Every diagram is proportional to the following integral
\begin{eqnarray}
{1\over  M_4^2}\sum_n\,\int\, \frac{d^4k}{(2\pi)^4}\,\frac{k^2}{k^2+
m_n^2}&=&{1\over  M_4^2}\sum_n\,\int\, \frac{d^4k}{(2\pi)^4}\,k^2\,
\int_0^\infty\, ds\, e^{-s(k^2+m_n^2)}\nonumber\\
&=& {1\over 8\pi^2 M_4^2}\,\sum_n\,\int_0^\infty\, \frac{ds}{s^3}\,
e^{-sm_n^2}.
\end{eqnarray}
As we saw for the gaugino case each diagram containing a gravitino
in Fig.~\ref{scalarfig} gives rise to the sum 
$\sum_{k} \beta_k \widetilde{\psi}_k\widetilde{\psi}_k$
where the coefficient $\beta_k$
depends on the location of the boundary chiral multiplet.
Thus, after subtracting the fermionic contributions from the bosonic
contributions the scalar mass-squared will be proportional to
\be
I_{\rm B}-I_{\rm F} =
{1\over 8\pi^2 M_4^2}\int_0^\infty{ds\over s^3}
\sum_n\left[ e^{-s n^2/R^2}-\beta_n e^{-s (n+\Delta)^2/R^2}\right]\, ,
\label{scaldiffsum}
\ee
where we have made use of the expression (\ref{eigval}). 
The infinite sum can be performed using standard techniques as explained
in the Appendix. For chiral matter on the brane $x_5=0$, the coefficient
$\beta_k=1$, and the sum of the diagrams in Fig.~\ref{scalarfig}
gives a contribution to the soft-breaking mass squared 
\be
\label{finalfinal}
m_\varphi^2 = - K^{-1}_{\varphi\varphi^\dagger}
K_{\varphi\varphi^\dagger}\, \frac{1}{8\pi^6 R^4 M_4^2}\,
\left\{\zeta(5)-\frac{1}{2}\left[
{\rm Li}_5\left(e^{-2\pi i\Delta}\right)+
{\rm Li}_5\left(e^{2\pi i\Delta}\right)\right]
\right\}\, ,
\ee
where $K(\varphi,\varphi^\dagger)=-3\,{\rm ln}\left[-\Omega(\varphi,
\varphi^\dagger)/3\right]$ is the K\"{a}hler potential and the
term $K^{-1}_{\varphi\varphi^\dagger}
K_{\varphi\varphi^\dagger}$ comes from the wave-function renormalization.
The supersymmetry-breaking contributions only come from the gravitino
diagrams, Fig.~\ref{scalarfig}(a) and (b).
The fact that the sign of $m_\varphi^2$ is negative can be 
easily understood by working in the rescaled Weyl basis. In that basis
loops involving gravitons do not give any contribution to the scalar masses 
which can only be induced by gravitino loops. The latter 
carry a negative sign because of their fermionic nature.

The one-loop scalar mass-squared induced by the
gravitational transmission of supersymmetry breaking is negative 
and diagonal in flavour space provided that the matter metric is diagonal. 
However, introducing other 
moduli fields $z$ in the bulk with gravitational strength coupling to the
boundary fields, one can get similar contributions to (\ref{finalfinal})
and make the total scalar  mass-squared positive~\cite{aq}. In such a case  
the K\"{a}hler potential will be a function of all scalar fields, 
$K=K(\varphi,\varphi^\dagger,z,z^\dagger)$, and different moduli 
dependence for the various scalar fields may create potentially 
dangerous flavour-changing neutral currents. 

Let us now take  two different and physically interesting 
limits of the expression (\ref{finalfinal}).
Consider first the case where the chiral multiplet is on the brane
at $x_5=0$.
If the absolute value of the superpotential $\left|W_\pi\right|$ is 
much smaller than $M_5^3$,
($\left|W_\pi\right|\ll M_5^3$), then we have seen that the function 
$\Delta$ is well approximated by  $\frac{W_\pi}{2\pi M_5^3}$.
Using the fact that for $\Delta\rightarrow 0$ we have
\be
\frac{1}{2}\left[{\rm Li}_5\left(e^{-2\pi i\Delta}\right)+
{\rm Li}_5\left(e^{2\pi i\Delta}\right)\right]
\simeq \zeta(5)-2\pi^2 \zeta(3) \Delta^2 + {\cal O}(\Delta^4)~,
\ee
we find that the leading contribution from the diagrams in 
Fig.~\ref{scalarfig} to the soft breaking scalar mass is proportional to 
\begin{equation}
\label{scalmass0}
m^2_\varphi\propto\left(
\frac{W_\pi}{R^2 M_4 M_5^3}\right)^2\sim 
\left(\frac{1}{R M_4}{\cal M}_{3/2}^{(0)}\right)^2~,
\end{equation}
where  we have used (\ref{grav0}). 
In this limit we see that the scalar mass 
acquires a suppression factor $(R M_4)^{-1}$
relative to the gravitino mass. When $RM_4 \sim 10$ then the scalar mass
will be comparable to the anomaly-mediated contribution~\cite{rs}.
The result (\ref{scalmass0}) can be
easily understood by noting that in the effective 4D supergravity 
there are no quadratic divergences.
The (divergent) one-loop scalar mass squared result 
in four-dimensions would be 
\be
m_\varphi^2 \sim\frac{m_{3/2}^2}{M_4^2}\,\int^\Lambda\, 
\frac{d^4p}{p^2+m_{3/2}^2}\sim
\frac{m_{3/2}^2}{M_4^2}\,\Lambda^2, 
\ee
where $\Lambda$ is the ultraviolet cutoff. However, in the brane world 
scenario the gravitino contribution to the scalar masses is the sum
of a pure 5D-divergent piece (cancelled by the graviton contribution) and a 
finite piece 
\be
m_\varphi^2 \sim\frac{m_{3/2}^2}{M_4^2}\,\int^{1/R}\, 
\frac{d^4p}{p^2+m_{3/2}^2}\sim
\frac{m_{3/2}^2}{M_4^2}\,\frac{1}{R^2}\, ,
\ee
where -- since the gravitino interacting with the chiral matter at
$x_5=0$ has to travel a distance at least as large as the radius of
compactification to probe supersymmetry breaking at $x_5=\pi R$ --
the loop integral is over 
gravitino  momenta satisfying 
$p < R^{-1}$ (gravitino wavelengths larger than $R$). 
Loop momenta 
$p > R^{-1}$ 
are not sensitive to the supersymmetry
breaking effects on the brane at $x_5=\pi R$. Thus, the effective ultraviolet
cutoff is provided by the interbrane distance and substituting
$\Lambda =1/R$ reproduces the result (\ref{scalmass0}).
This situation resembles what happens
in 4D theories at finite temperature where the ultraviolet
cutoff is represented by the temperature $T$.
In the 
imaginary time formalism
4D loop integrals become integrals over the three
spatial momenta and a sum over the so-called Matsubara frequencies
and the one-loop contributions to the mass squared of an interacting 
scalar field can be split into the 
usual zero temperature 4D divergent piece plus
a finite temperature dependent piece. The finiteness
is due to the fact that particles in the plasma with wavelengths
smaller than $T^{-1}$ (or momenta larger than $T$) have
exponentially suppressed abundances.

From Eq. (\ref{scalmass0}) we find that for ${\cal M}_{3/2}^{(0)}R^{-1}\sim
\left(10^{11}\,\,{\rm GeV}\right)^2$, the soft scalar masses are of
order of the TeV scale. For instance, if $W_\pi/M_5^3\sim 10^{-2}$, 
we get $M_5\sim 5\times 10^{16}$ GeV and $R^{-1}\sim 10^{12}$ GeV.

If the absolute value of the superpotential $\left|W_\pi\right|$ is 
much larger than $M_5^3$, $\left|W_\pi\right|\gg M_5^3$, then we have 
seen that the function $\Delta$ is well approximated by  $1/2$, and
the polylogarithmn functions can be expanded as
\be
\frac{1}{2}\left[{\rm Li}_5\left(e^{-2\pi i\Delta}\right)+
{\rm Li}_5\left(e^{2\pi i\Delta}\right)\right]
\simeq -\frac{15}{16}\zeta(5) 
+ {\cal O}\left((\Delta-\frac{1}{2})^2\right)~.
\ee
In such a case we find that for large values of $\left|W_\pi\right|$ 
the integral
${\cal I}=\frac{93}{32\pi^5} \zeta(5)$, as defined in the Appendix. 
This reproduces the value found in Ref.~\cite{aq}
where supersymmetry was broken by boundary conditions
via the Scherk-Schwarz mechanism. This does not come
as a surprise, though. As we have already pointed out,
the gravitino spectrum in the limit $\left|W_\pi\right|\gg M_5^3$,
is exactly the same as obtained in the Scherk-Schwarz supersymmetry 
breaking mechanism.  The soft breaking scalar mass is therefore 
proportional to 
\begin{equation}
\label{scalmassinf}
m_\varphi \propto
\frac{1}{R^2 M_4}\simeq
\frac{\left({\cal M}_{3/2}^{(0)}\right)^2}{M_4}~,
\end{equation} 
where we have used (\ref{gravinf}) and there is an $R M_4$
suppression. Again this result can be understood by noting that the 
effective cutoff for the brane-world scenario is $\Lambda = 1/R$.
If $R^{-1}\sim 10^{11}$ GeV, we find $m_\varphi$ of the order of 
TeV.

It is also instructive to discuss the alternate possibility
when the chiral matter
is located on the same brane where supersymmetry breaking occurs
($\beta_n=\cos^2(n+\Delta)\pi$). In this case since loop momenta are on
the same brane that has a nonzero superpotential the interbrane distance 
$R$ no longer plays any role. As happens
in the four-dimensional case, scalar masses should be ultraviolet
sensitive to the cutoff $\Lambda$. This is because gravitinos are now repelled
from the brane and the cancellation of the five-dimensional divergences
no longer takes place. This can be explicitly seen by performing 
the integral (\ref{scaldiffsum}) and the details can be found in the Appendix.
If the supersymmetry breaking parameter $|W_\pi|$ is much smaller
than $M_5^3$ one finds that  $m_\varphi\sim
{\cal M}_{3/2}^{(0)}(\Lambda/M_4) (\Lambda R)$ which is the usual
4D dimensional divergences increased by the number of Kaluza-Klein states
$(\Lambda R)$ excited up to the cutoff scale. If $|W_\pi|$ is much larger
than $M_5^3$, one finds $m_\varphi\sim \sqrt{\Lambda R} (\Lambda^2/M_4) $ 
which reflects the absence of the cancellation of 
the pure five-dimensional divergences.

\subsection{The 5D calculation}

The communication of supersymmetry breaking to the boundary matter
fields can also be obtained using the five-dimensional propagator for
the gravitino field. In order to calculate the 5D propagator it is 
simplest to Fourier transform the four-dimensional spatial coordinates, while
leaving the fifth spatial coordinate $x_5$ explicit~\cite{ahgs}. 
The propagator of the massless gravitino in 5D can be written in the form
\be
    G_{\mu\nu}(k,x_5) = G(k,x_5)\left(g_{\mu\nu}-\frac{1}{3}
      \gamma_\mu\gamma_\nu\right)\equiv G(k,x_5)P_{\mu\nu}~,
\ee
where we have omitted the longitudinal parts because they do
not give any contribution to the gaugino and scalar masses.
After the addition of the boundary mass term this reproduces the
propagator of the gravitino transverse degrees of freedom. 
To evaluate $G(k,x_5)$ in the case of supersymmetry breaking we
have to solve, starting from Eqs. (\ref{one}) and (\ref{two}) and
their conjugates, the equations for the Green's function $G_{1,2}$. 
For simplicity we assume that there is a boundary mass term, $m\equiv 
W_\pi/ M_5^3$, at $x_5=\pi R$. Following the procedure developed 
in Refs.~\cite{ahhnsw,kapwein}
we first solve for the gravitino propagator in infinite space. It reads
\begin{eqnarray}
\label{infprop}
    {\widetilde G}_1(k_E,x_5)&=& \frac{-i\sigma\cdot k_E}{2k_E} 
    \left(e^{-k_E |x_5|}-\frac{m^2}{4+m^2} e^{-k_E \pi R-k_E|x_5-\pi R|}
   \right) + \frac{i}{2k_E}
    \partial_5 e^{-k_E |x_5|}~\nonumber\\
    &-&\frac{m}{4+m^2} e^{-k_E \pi R-k_E|x_5-\pi R|}~,
\end{eqnarray}
where we have analytically continued the propagator to be a function
of the Euclidean four-momentum $k_E$. Let us consider an arbitrary
point $x_5$ in the interval $[0,\pi R]$. The amplitude to propagate from
$x_5$ to $x_5$ is simply ${\widetilde G}_1(k_E,x_5)$. 
But since our space is compact
and not infinite, the point $x_5+2\pi n R$ where $n$ is an integer, 
is equivalent to $x_5$ and we need to sum over all the contributions, 
${\widetilde G}_1(k_E,x_5+2\pi n R)$. This sum can be easily performed 
and leads to the result for a compact space with $0\leq x_5 \leq \pi R$
\begin{eqnarray}
    G_1^{(\Delta)}(k_E,x_5)&=&\frac{-i\sigma\cdot k_E}{2k_E\sinh(\pi k_E R)} 
      \left[\cosh[k_E(\pi R- x_5)]-\frac{m^2 \cosh(k_E x_5)\cosh(\pi k_E R)}
{4\sinh^2(\pi k_E R)+m^2\cosh^2(\pi k_E R)}\right]\nonumber\\
   &&\qquad\qquad\qquad-\frac{m \cosh(\pi k_E R)}{4\sinh^2(\pi k_E R)
    +m^2\cosh^2(\pi k_E R)}\nonumber~,\\
   &=&\frac{-i\sigma\cdot k_E}{2k_E} \left[
  \frac{\sinh[k_E(2\pi R - x_5)] + \cos(2\pi\Delta)\sinh(k_E x_5)}
  {\cosh(2\pi k_E R) -\cos(2\pi\Delta)}\right]\nonumber\\
   &&\qquad\qquad\qquad\qquad -\frac{\cosh(k_E x_5)\sin(2\pi\Delta)}
  {2[\cosh(2\pi k_E R)-\cos(2\pi\Delta)]}~,
\end{eqnarray}
where $m=2\tan\Delta\pi$, using the relation (\ref{del}) for $W_0=0$.
Similarly, one can follow the same procedure to obtain for
$0\leq x_5 < \pi R$
\begin{eqnarray}
    G_2^{(\Delta)}(k_E,x_5)&=&\frac{-i\bar\sigma\cdot k_E}
       {2k_E\sinh(\pi k_E R)} 
      \left[\cosh[k_E(\pi R- x_5)]-\frac{2 m\sinh(k_E x_5)}
   {4\sinh^2(\pi k_E R)+m^2\cosh^2(\pi k_E R)}\right]\nonumber\\
   &&\qquad\qquad\qquad-\frac{m^2\sinh(k_E x_5)\coth(\pi k_E R)}
   {2[4\sinh^2(\pi k_E R)+m^2\cosh^2(\pi k_E R)]}~.
\end{eqnarray}
Notice that we have not written 
the $\gamma_5$ terms since they give no contribution to the masses.

Let us consider the case where matter is on the brane at $x_5=0$. 
We find that the $G_2$ propagator is particularly simple, 
\be
   G_2^{(\Delta)}(k_E,0) =\frac{-i\bar\sigma\cdot k_E}{2k_E} 
   \coth(\pi k_E R)~,
\ee
and does not depend on the supersymmetry breaking parameter, $\Delta$.
However, the $G_1$ propagator becomes
\be
  \label{gravprop}
   G_1^{(\Delta)}(k_E,0) =
   \frac{-i(\sigma\cdot k_E)/k_E\, \sinh(2\pi k_E R)
   -\sin(2\pi\Delta)}{2[\cosh(2\pi k_E R)-\cos(2\pi\Delta)]}~.
\ee
In the supersymmetric limit, $\Delta\rightarrow 0$ we recover the usual 
5D propagator
\be
 G_1^{(0)}(k_E,0)= \frac{-i\sigma\cdot k_E}{2k_E}\coth(\pi k_E R)~,
\ee
while in the maximally supersymmetry-breaking limit, $\Delta
\rightarrow 1/2$ (or Scherk-Schwarz limit), we obtain 
\be
  G_1^{(1/2)}(k_E,0)= \frac{-i\sigma\cdot k_E}{2k_E}\tanh(\pi k_E R)~.
\ee
The propagator (\ref{gravprop}) thus continuously interpolates between
these two limits. Notice also
that after analytically continuing the momentum back to four-dimensional
Minkowski space the poles of the propagator (\ref{gravprop}) occur at 
the values $ik_E\equiv k_4=(n+\Delta)/R$, in complete agreement
with (\ref{eigval}).
It is also interesting to consider the four-dimensional limit
$k_E R\ll 1$. In this limit the propagator (\ref{gravprop}) becomes
\be
    G_1^{(\Delta)}(k_E,0) = \frac{1}{2 \pi R} 
   \frac{-i\sigma\cdot k_E - \Delta/R}
     {k_E^2+\Delta^2/R^2}~,
\ee
where we have also taken the limit $\Delta\rightarrow 0$.

When matter is located on the brane at $x_5=\pi R$ we obtain
\be
 \label{gravpropPiR}
   G_1^{(\Delta)}(k_E,\pi R) = \frac{-i\sigma\cdot k_E}{k_E} 
   \left[\frac{\cos^2(\pi\Delta)\sinh(\pi k_E R)}
  {\cosh(2\pi k_E R) -\cos(2\pi\Delta)}\right]
   -\frac{\cosh(\pi k_E R)\sin(2\pi\Delta)}
  {2[\cosh(2\pi k_E R)-\cos(2\pi\Delta)]}~,
\ee
and we see that there is an extra cosine factor which can be thought of
as being due to the wavefunction of the gravitino at $x_5=\pi R$.

It is now straightforward to obtain the one-loop contributions
to the boundary matter. Consider first the case of the gaugino mass
on the boundary $x_5=0$. One finds that the 
contribution to the gaugino mass from the Feynman diagrams
in Fig.~\ref{gauginofig}(a) is simply
\begin{eqnarray}
   &&\frac{-i}{4M_5^3}\int \frac{d^4 k}{(2\pi)^4}\,\frac{\sin(2\pi\Delta)}
  {2[\cosh(2\pi k_E R)-\cos(2\pi\Delta)]}\nonumber \\
    &=& \frac{-i}{64\pi^2}\frac{1}{(\pi R)^4 M_5^3} \int_0^\infty dx\, x^3
       \frac{\sin(2\pi\Delta)} {\cosh(2x)-\cos(2\pi\Delta)}\nonumber\\
    &=& \frac{1}{64\pi^5}\frac{3}{8 M_4^2 R^3}
     \left[{\rm Li}_4(e^{-2\pi i\Delta}) 
        - {\rm Li}_4(e^{2\pi i\Delta})\right]~,
\end{eqnarray}
and similarly for Fig.~\ref{gauginofig}(b)
\be
    -\frac{1}{64\pi^5}\frac{3}{8 M_4^2 R^3}
     \left[{\rm Li}_4(e^{-2\pi i\Delta}) 
        - {\rm Li}_4(e^{2\pi i\Delta})\right]~.
\ee
Thus, as expected we see that the sum of the one-loop contributions
to the gaugino mass is zero. 
This also agrees with the result in Ref.~\cite{barbieri}, 
which found that there are no gravitational
radiative corrections when the theory has a cutoff.
In the 5D theory the effective cutoff is $\sim 1/R$, 
and we obtain a similar result. Notice also that each separate 
contribution to the gaugino mass vanishes identically for 
$\Delta=0, 1/2$. In the Kaluza-Klein picture this
can be easily seen since for $\Delta =0,1/2$ the 
Kaluza-Klein mass spectrum is symmetric about zero mass 
and leads to a vanishing sum (\ref{summ}).
Similarly, one obtains the same results for matter
on the brane at $x_5=\pi R$, since the only relevant 
difference is the cosh factor in the propagator (\ref{gravpropPiR}).

In the case of the boundary scalar fields located at $x_5=0$, 
and using the Lagrangian 
(\ref{matter}) the contribution to the scalar mass-squared from the 
gravitino loops is given by
\be
\label{gravcont}
    \frac{-1}{6 M_5^3}\int \frac{d^4 k}{(2\pi)^4}\, 
      \epsilon^{\mu\nu\rho\sigma}
        {\rm Tr}\left[\gamma_5 \gamma_\nu k_\rho
            G_\Delta(k,0)P_{\mu\sigma} \right] ~,
\ee
where $G_{\Delta}(k,0)$ contains 
$G_1^{(\Delta)}$ and ${\bar G}_1^{(\Delta)}$.
There are also contributions from the graviton and graviphoton, which 
in the limit of zero supersymmetry breaking must cancel the gravitino 
contribution (\ref{gravcont}). Thus, the total contribution to the scalar 
mass-squared is given by
\begin{eqnarray}
\label{5dint}
  &&  \frac{1}{6M_5^3} \int \frac{d^4 k}{(2\pi)^4}\, (-8) k_E^2\,
    \frac{1}{2 k_E}\left[\coth(\pi k_E R) - \frac{\sinh(2\pi k_E R)}
       {\cosh(2\pi k_E R) - \cos(2\pi \Delta)}\right]\nonumber\\
    &=& \frac{-1}{12\pi^2 (\pi R)^5 M_5^3} \int_0^\infty dx\, x^4
       \left[\coth(x) - \frac{\sinh(2x)}{\cosh(2x)
         -\cos(2\pi \Delta)}\right]\nonumber\\
    &=& \frac{-1}{8\pi^6 R^4 M_4^2}
\left\{\zeta(5)-\frac{1}{2}\left[
{\rm Li}_5\left(e^{-2\pi i\Delta}\right)+
{\rm Li}_5\left(e^{2\pi i\Delta}\right)\right]\right\}~.
\end{eqnarray}
This agrees with the result obtained from the Kaluza-Klein sum 
(\ref{finalfinal}).
Notice that we obtain a finite result because the leading divergences
cancel in the integral (\ref{5dint}), 
and the remaining part is exponentially suppressed.

When matter is located on the brane at $x_5=\pi R$ this cancellation
no longer happens because the gravitino propagator (\ref{gravpropPiR})
contains a cosine factor which depends on $\Delta$. Only in the 
supersymmetric limit $(\Delta=0)$ does the
cancellation in (\ref{5dint}) occur. 
This is just the 5D interpretation of the result 
(\ref{infKK}) that we obtained earlier from the Kaluza-Klein summation.

\section{Conclusions and discussions}

In this paper we have considered a supersymmetric 
five-dimensional brane-world scenario
where the fifth dimension is compactified on
$S^1/Z_2$. In our set-up chiral matter and gauge fields are restricted 
to live on the boundaries while gravity propagates in the
bulk. We have assumed that supersymmetry is broken at the orbifold fixed points
and that supersymmetry breaking is parametrized by a constant 
boundary superpotential. The bulk gravitino mass spectrum is consequently
shifted relative to the bosonic bulk supergravity fields. Integrating
out the supergravity auxiliary fields allows one to derive the couplings
between the boundary chiral matter or gauge fields and the bulk supergravity
fields. If chiral matter or gauge fields live on a brane different 
from the one where supersymmetry breaking occurs, the latter is communicated
to our observable world by gravitational interactions. 

We have computed the contribution to the soft supersymmetry breaking 
scalar and gaugino masses for generic values of the brane superpotential 
and the radius of compactification. The one-loop computation of the
mass splittings generically provides a hierarchy of soft masses at the 
TeV scale with nonvanishing
scalar masses and zero gaugino masses.

The nonvanishing superpotential on the boundaries
and the relative shift in the bulk gravitino mass spectrum parametrize
various possible sources of supersymmetry breaking at the fixed-points
coming from Fayet-Iliopoulos $D$- and $F$-terms inducing nonvanishing
brane superpotentials to keep the branes tensionless.
In this paper we have assumed that these terms are coupled to the bulk 
only through supergravity. In addition for certain values of the
radius $R$ the anomaly-mediated contributions of scalar masses 
will be of roughly the same order. It may be that a combination of 
brane supersymmetry breaking and the anomaly-mediated contribution can 
solve the well-known slepton mass problem. On the other hand 
since the one-loop gaugino masses vanish, the anomaly-mediated gaugino 
mass contributions will dominate.

In deriving the results of this paper we have worked with the
components of the supermultiplets. In principle the same results can also 
be more simply derived using the $N=1$ superfield calculus, such as that 
considered in Refs.~\cite{n1cal} for the case of bulk vector fields. 
This would require writing the $N=2$ supergravity Lagrangian in terms 
of $N=1$ superfields, and then repeating the procedure already done for 
the bulk gauge fields.

One should note that the supersymmetry breaking mechanism
respects the tensionlessness of the branes even though we have added
matter on the branes. This is because the Goldstino is a bulk field. 
Of course, a brane tension may eventually be generated due to gauge 
symmetry breaking (or even other forms of brane supersymmetry breaking). 
This may require considering more general warped bulk backgrounds. 
Since we have used the tensor multiplet it is straightforward
to generalize the procedure used here for warped geometries. In 
particular one could obtain the gravitational analogue of the 
warped soft mass spectrum calculated in Ref.~\cite{gp}.

Finally, even if the branes remain tensionless, the vacuum 
energy is nonzero~\cite{bfz} and the radius is not stabilized. 
Thus, one is likely to require the addition of further fields in the 
bulk in order to stabilize the radius~\cite{ls,pp}. These fields will
arise when one embeds the brane-world setup
in a more fundamental theory, such as string theory, and deserves 
further study.

\section*{Acknowledgements}
We wish to thank I. Antoniadis, A. Brignole, 
F. Feruglio, Z. Lalak, A. Pomarol,
M. Quiros, R. Rattazzi, M. Zucker and F. Zwirner for
useful discussions. T.G. thanks the  
University of Padova Theory Group
and A.R. thanks the CERN Theory Group for hospitality 
where part of this work was done.
The work of T.G. is supported by the Swiss National 
Science Research Fund (FNRS) contract no. 21-55560.98.

\newpage
\section*{Appendix}

We provide the details on evaluating the infinite Kaluza-Klein sum
 (\ref{scaldiffsum}). The key is to introduce
the $\Theta$-functions defined as
\be
\Theta_\beta^\alpha(\tau)=\sum_{n=-\infty}^{\infty}\,
e^{2\pi i n \beta}\, e^{\pi i \tau(n +\alpha)^2}\, ,
\ee
which obey the Poisson resummation formula
\be
\label{poisson}
\Theta_\beta^\alpha(-1/\tau)=\sqrt{-i\tau}\, 
e^{-2\pi i \alpha \beta}\, \Theta_\alpha^{-\beta}(\tau)~.
\ee

Consider the expression (\ref{scaldiffsum}) with $\beta_k=1$ which
can be rewritten in the form
\begin{eqnarray}
I_{\rm B}-I_{\rm F} &=&
{1\over 8\pi^2 M_4^2}\int_0^\infty{ds\over s^3}
\left[ \Theta_0^0\left({is\over\pi R^2}\right)-
\Theta_0^\Delta\left({is\over\pi R^2}\right)\right]\nonumber~,\\
&=&{1\over 8\pi^2 M_4^2}\int_0^\infty{ds\over s^3}
\left({\pi R^2\over s}\right)^{1/2}
\left[ \Theta_0^0\left({i\pi R^2\over s}\right)-
\Theta_\Delta^0\left({i\pi R^2\over s}\right)\right]~,
\label{dint}
\end{eqnarray}
where we have used (\ref{poisson}).
After redefining the integration variable to be $y= R^2/s$, we obtain
\begin{eqnarray}
\label{intds}
I_{\rm B}-I_{\rm F}
&=&\frac{1}{8\pi\sqrt{\pi} R^4 M_4^2}\int_0^\infty dy\, y^{3/2}
\left[ \Theta_0^0\left(i \pi y\right)-
\Theta_\Delta^0\left(i\pi y\right)\right]~,\nonumber\\
&=&\frac{1}{8\pi\sqrt{\pi} R^4 M_4^2}\int_0^\infty dy\, y^{3/2}
\sum_{n=-\infty}^{\infty}\, \left(1- e^{2\pi i n \Delta}\right)
e^{-\pi^2 n^2 y}~,\nonumber\\
&\equiv&
\frac{1}{8\pi R^4 M_4^2}\,{\cal I}\, ,
\end{eqnarray}
where we have defined
\be 
{\cal I}=\frac{1}{\sqrt{\pi}}\int_0^\infty\, dy\, y^{3/2}
\sum_{n=-\infty}^{\infty}\, 
\left(1- e^{2\pi i n \Delta}\right) e^{-\pi^2 n^2 y}~.
\ee
When $n$ is nonzero the exponential factor guarantees that each
term in the sum has a finite integral. However, notice that the
potentially dangerous $n=0$ term vanishes. Thus, simplifying the 
infinite sum gives
\be
\sum_{n=-\infty}^{\infty}\, \left(1- e^{2\pi i n \Delta}\right)
e^{-\pi^2 n^2 y} =
4\,\sum_{n=1}^{\infty}\,\sin^2(\pi n \Delta)e^{-\pi^2 n^2 y}~.
\ee
so that performing the $y$ integration and summing up
the finite integral pieces gives the finite answer
\begin{eqnarray}
{\cal I}&=& 
\frac{3}{\pi^5}\, \sum_{n=1}^{\infty}\,
\sin^2(\pi n \Delta)\frac{1}{n^5}~, \nonumber\\
&=& 
\frac{3}{2 \pi^5}\,\left\{\zeta(5)-\frac{1}{2}\left[
{\rm Li}_5\left(e^{-2\pi i\Delta}\right)+
{\rm Li}_5\left(e^{2\pi i\Delta}\right)\right]
\right\}\, ,
\end{eqnarray}
where ${\rm Li}_{k}(x)=\sum _{n=1}^{\infty}\,\frac{x^n}{n^k}$ are the
polylogarithm functions. Thus, substituting the value of ${\cal I}$ back
into the expression (\ref{intds}) gives the final result
\be
\label{final}
I_{\rm B}-I_{\rm F}=
\frac{3}{2 \pi^5}\,\frac{1}{8\pi R^4 M_4^2}\,
\left\{\zeta(5)-\frac{1}{2}\left[
{\rm Li}_5\left(e^{-2\pi i\Delta}\right)+
{\rm Li}_5\left(e^{2\pi i\Delta}\right)\right]
\right\}~,
\ee
which leads to the result (\ref{finalfinal}).

In the case where the chiral matter lives on the same brane as the
supersymmetry breaking, the evaluation of the infinite sum is only
slightly more involved because of the presence of the 
factor $\beta_n=\cos^2(n+\Delta)\pi$. In this case (\ref{scaldiffsum})
becomes
\be
\label{intsum1}
I_{\rm B}-I_{\rm F} =
{1\over 8\pi^2 M_4^2}\int_0^\infty{ds\over s^3}
\,\sum_{n=-\infty}^{\infty}\,
\left[ e^{-s n^2/R^2}-\cos^2\left[\left(n+\Delta\right)\pi\right]
e^{-s (n+\Delta)^2/R^2}\right]\, ,
\ee
and the only complication involves the evaluation of the infinite
sum with the cosine factor.
Using the expansion $\cos\left[\left(n+\Delta\right)\pi\right]=
\left[e^{i\pi(n+\Delta)}+e^{-i\pi(n+\Delta)}\right]/2$, one can easily 
show that
\be
\sum_{n=-\infty}^{\infty}\,
\cos^2\left[\left(n+\Delta\right)\pi\right]\,e^{-s (n+\Delta)^2/R^2}
= \cos^2(\Delta\pi) \sqrt{\frac{\pi R^2}{s}}
\Theta_\Delta^0(i \pi R^2/s)~.
\ee
Thus, after changing variables to $y=R^2/s$
the infinite sum (\ref{intsum1}) can be rewritten in the form
\begin{eqnarray}
\label{intdspi}
I_{\rm B}-I_{\rm F}
&=&\frac{1}{8\pi\sqrt{\pi} R^4 M_4^2}\int_0^\infty dy\, y^{3/2}
\left[ \Theta_0^0\left(i \pi y\right)-\cos^2(\Delta\pi)
\Theta_\Delta^0\left(i\pi y\right)\right]~,\nonumber\\
&=&\frac{1}{8\pi\sqrt{\pi} R^4 M_4^2}\int_0^\infty dy\, y^{3/2}
\sum_{n=-\infty}^{\infty}\, \left(1- \cos^2(\Delta\pi)
e^{2\pi i n \Delta}\right) e^{-\pi^2 n^2 y}~,\nonumber\\
&\equiv& \frac{1}{8\pi R^4 M_4^2}\,{\cal I_\pi}\, ,
\end{eqnarray}
where we have defined
\be 
{\cal I_\pi}=\frac{1}{\sqrt{\pi}}\int_0^\infty\, dy\, y^{3/2} 
\sum_{n=-\infty}^{\infty}\, 
\left(1- \cos^2(\Delta\pi) e^{2\pi i n \Delta}\right) e^{-\pi^2 n^2 y}~.
\ee
Notice now that the $n=0$ term in the sum no longer vanishes, and
there is a divergent piece in ${\cal I_\pi}$. Thus, simplifying the infinite 
sum gives
\be
\label{infKK}
\sum_{n=-\infty}^{\infty}\, \left[1- \cos^2(\Delta\pi)
e^{2\pi i n \Delta}\right] e^{-\pi^2 n^2 y^2} =
\sin^2\Delta\pi + 2\,\sum_{n=1}^{\infty}\,\left[1-\cos^2(\pi \Delta)
\cos(2\pi n \Delta)\right]\, e^{-\pi^2 n^2 y^2}~.
\ee
Only in the limit $\Delta =0$ does the divergent piece from the $n=0$ term
vanish. This makes sense since there is now no superpotential breaking term.


\newpage

\begin{thebibliography}{99}

\bibitem{anton}
I.~Antoniadis,
Phys.\ Lett.\ B {\bf 246} (1990) 377.

\bibitem{hw}
P.~Horava and E.~Witten,
Nucl.\ Phys.\ B {\bf 460} (1996) 506
[hep-th/9510209];
P.~Horava and E.~Witten,
Nucl.\ Phys.\ B {\bf 475} (1996) 94
[hep-th/9603142];
P.~Horava,
Phys.\ Rev.\ D {\bf 54} (1996) 7561
[hep-th/9608019].

\bibitem{aq}
I.~Antoniadis and M.~Quiros,
Nucl.\ Phys.\ B {\bf 505} (1997) 109
[hep-th/9705037].

\bibitem{nilles}
H.~P.~Nilles, M.~Olechowski and M.~Yamaguchi,
Phys.\ Lett.\ B {\bf 415} (1997) 24
[hep-th/9707143];
H.~P.~Nilles, M.~Olechowski and M.~Yamaguchi,
Nucl.\ Phys.\ B {\bf 530}, 43 (1998)
[hep-th/9801030];
H.~P.~Nilles,
hep-ph/0004064.

\bibitem{mirpes}
E.~A.~Mirabelli and M.~E.~Peskin,
Phys.\ Rev.\ D 58 (1998) 065002 [hep-th/9712214].

\bibitem{rs}
L.~Randall and R.~Sundrum,
Nucl.\ Phys.\ B {\bf 557} (1999) 79
[hep-th/9810155].

\bibitem{adpq}
I.~Antoniadis, S.~Dimopoulos, A.~Pomarol and M.~Quiros,
Nucl.\ Phys.\ B {\bf 544} (1999) 503
[hep-ph/9810410];
A.~Delgado, A.~Pomarol and M.~Quiros,
Phys.\ Rev.\ D {\bf 60} (1999) 095008
[hep-ph/9812489].

\bibitem{gm}
D.~E.~Kaplan, G.~D.~Kribs and M.~Schmaltz,
Phys.\ Rev.\ D {\bf 62} (2000) 035010
[hep-ph/9911293];
Z.~Chacko, M.~A.~Luty, A.~E.~Nelson and E.~Ponton,
JHEP {\bf 0001} (2000) 003
[hep-ph/9911323].

\bibitem{gp}
T.~Gherghetta and A.~Pomarol,
Nucl.\ Phys.\ B {\bf 602} (2001) 3
[hep-ph/0012378].

\bibitem{bfz} 
J.~A.~Bagger, F.~Feruglio and F.~Zwirner,
hep-th/0107128;
J.~Bagger, F.~Feruglio and F.~Zwirner,
hep-th/0108010.

\bibitem{zucker}
M.~Zucker,
Nucl.\ Phys.\ B {\bf 570} (2000) 267
[hep-th/9907082];
M.~Zucker,
JHEP {\bf 0008} (2000) 016
[hep-th/9909144];
M.~Zucker,
Phys.\ Rev.\ D {\bf 64} (2001) 024024
[hep-th/0009083].

\bibitem{kugo}
T.~Kugo and K.~Ohashi,
Prog.\ Theor.\ Phys.\  {\bf 104} (2000) 835
[hep-ph/0006231];
T.~Kugo and K.~Ohashi,
Prog.\ Theor.\ Phys.\  {\bf 105} (2001) 323
[hep-ph/0010288];
T.~Fujita, T.~Kugo and K.~Ohashi,
hep-th/0106051.

\bibitem{ss}
J.~Scherk and J.~H.~Schwarz,
Phys.\ Lett.\ B {\bf 82} (1979) 60;
J.~Scherk and J.~H.~Schwarz,
Nucl.\ Phys.\ B {\bf 153} (1979) 61;
E.~Cremmer, J.~Scherk and J.~H.~Schwarz,
Phys.\ Lett.\ B {\bf 84} (1979) 83.

\bibitem{glmr}
G.~F.~Giudice, M.~A.~Luty, H.~Murayama and R.~Rattazzi,
JHEP {\bf 9812} (1998) 027
[hep-ph/9810442].

\bibitem{pure5d}
E.~Cremmer, in {\em Superspace and Supergravity}, S.W.~Hawking and 
M.~Rocek eds., Cambridge University Press, 1981, pp.267--282;
A.~H.~Chamseddine and H.~Nicolai,
Phys.\ Lett.\ B {\bf 96} (1980) 89.

\bibitem{sw}
M.~F.~Sohnius and P.~C.~West,
Nucl.\ Phys.\ B {\bf 216} (1983) 100.

\bibitem{lalak} 
A.~Falkowski, Z.~Lalak and S.~Pokorski,
hep-th/0102145.

\bibitem{mno}
K.~A.~Meissner, H.~P.~Nilles and M.~Olechowski,
Nucl.\ Phys.\ B {\bf 561} (1999) 30
[hep-th/9905139].

\bibitem{ddg}
K.~R.~Dienes, E.~Dudas and T.~Gherghetta,
Phys.\ Rev.\ D {\bf 62} (2000) 105023
[hep-ph/9912455].

\bibitem{barbieri}
R.~Barbieri, L.~Girardello and A.~Masiero,
Phys.\ Lett.\ B {\bf 127} (1983) 429;
P.~Binetruy, S.~Dawson and I.~Hinchliffe,
Phys.\ Rev.\ D {\bf 35} (1987) 2215.

\bibitem{nibbelink}  
S.~Groot Nibbelink,
hep-th/0108185.

\bibitem{ls} 
M.~A.~Luty and R.~Sundrum,
Phys.\ Rev.\ D {\bf 62} (2000) 035008
[hep-th/9910202].

\bibitem{ahgs}
N.~Arkani-Hamed, Y.~Grossman and M.~Schmaltz,
Phys.\ Rev.\ D {\bf 61} (2000) 115004
[hep-ph/9909411].

\bibitem{ahhnsw}
N.~Arkani-Hamed, L.~J.~Hall, Y.~Nomura, D.~R.~Smith and N.~Weiner,
Nucl.\ Phys.\ B {\bf 605} (2001) 81
[hep-ph/0102090].

\bibitem{kapwein}
D.~E.~Kaplan and N.~Weiner,
hep-ph/0108001.

\bibitem{n1cal}
N.~Arkani-Hamed, T.~Gregoire and J.~Wacker,
hep-th/0101233;
D.~Marti and A.~Pomarol,
hep-th/0106256.

\bibitem{pp}
E.~Ponton and E.~Poppitz,
JHEP {\bf 0106} (2001) 019
[hep-ph/0105021].


\end{thebibliography}
\end{document}